\documentclass[draft]{agujournal2019}
\usepackage{url} 
\usepackage{lineno}
\usepackage{amsmath}
\usepackage{amssymb}
\usepackage{tikz}
\usepackage{soul}

\draftfalse

\journalname{JGR: Planets}

\begin{document}

%
%


\title{Exploring Enceladus’s Interior Structure Using Electromagnetic Induction}

\authors{Alexander Grayver, Joachim Saur}

\affiliation{}{Institute of Geophysics and Meteorology, University of Cologne, Germany}

\correspondingauthor{A. Grayver}{agrayver@uni-koeln.de}

\begin{keypoints}
\item Global and local electromagnetic response functions are elaborated for orbiter and lander configurations on Enceladus.
\item Ice-shell thickness gradients induce 3-D magnetic anomalies that correlate with the shell structure and depend on ocean conductivity.
\item Broadband lander EM transfer function at periods $10^1 - 10^5$ s can constrain ocean conductivity, ocean thickness, and core properties.
\end{keypoints}

\begin{abstract}
Electromagnetic (EM) sounding can constrain the electrical structure of Enceladus and, in turn, the salinity of its ocean and the porosity, fluid content, and thermal state of its hydrothermally active core. Here, we assess the feasibility of EM sounding at Enceladus using both global (orbiter) and local (lander) EM induction transfer functions. We provide a physical framework for modeling EM induction for 1-D and 3-D subsurface conductivity models and discuss how transfer functions can be estimated from global or local measurements of the magnetic and electric fields.
We simulate 3-D induction effects arising from variations in ice-shell thickness. The magnitude of these effects in the magnetic field correlates with the ice-shell thickness at the surface and is strongly dependent on the ocean's conductivity. These magnetic variations, if observed, would favor a moderately to highly conductive ocean, providing lower bounds on salinity and volatile content. The absence of these effects indicates a thicker, more homogeneous ice shell and/or a lower-conductivity ocean. Given plausible magnitudes, a polar-orbiting mission with low-altitude measurements will be required to detect these effects.
In summary, an orbiter will constrain global ocean conductivity using long-period induction and possibly map the ice thickness variations. The detailed EM sounding of both the hydrosphere and the core can be achieved by a lander-based broadband EM sounding at periods $\approx 10^1-10^5$ s to probe ocean salinity and thickness, as well as core properties including porosity, fluid content, and temperature. 
\end{abstract}

\section*{Plain Language Summary}

Enceladus is an icy moon orbiting Saturn. There is strong evidence that it harbours a global liquid ocean, a porous and hydrothermally active core, and geysers at its south pole. Observations suggest its ice shell is thinner at the poles than at the equator. These characteristics make Enceladus a prime target for future space missions that aim to determine whether its ocean could be habitable. Answering that question depends, among other factors, on the still poorly constrained physical properties and structure of the moon's interior. Because we cannot directly access the ocean or core, we can use a remote-sensing technique such as electromagnetic (EM) sounding, in which the interior response to external EM forcing is related to electrical conductivity. Electrical conductivity is sensitive to key properties such as salinity and temperature inside Enceladus. This study presents a comprehensive analysis and simulation of EM responses for plausible interior structures and provides guidance for future mission observations aimed at maximizing scientific return.

\section{Introduction}

Enceladus possesses many key requirements for habitability \cite{glein2018geochemistry, choblet2022enceladus}, including the presence of liquid water, essential chemical elements, and a source of energy, making it a top target for future space missions. As part of the ESA Voyage 2050 program, Enceladus has been selected as the target of the next L4 class mission \cite{martins2024report}. In the most ambitious configuration, the mission aims to implement both an orbiter and a lander, each carrying a science payload. 

Among other geophysical techniques, electromagnetic (EM) sounding allows us to probe the electrical conductivity of the interior. Conductivity is highly sensitive to key properties, such as ocean salinity, as well as to fluid content and temperature within Enceladus's hydrothermally active, porous core. To enable EM sounding, measurements of the magnetic and possibly electric fields around and/or at the surface of Enceladus are required. However, in contrast to the Jovian system, where an inclined magnetic dipole axis of Jupiter drives strong EM induction effects in the Galilean moons at well charactaerized periods (e.g., \citeA{neubauer1999alfven, kivelson2000galileo, saur2010induced}), the axisymmetric structure of the Kronian magnetic field \cite{dougherty2018saturn, dougherty2018review} and the ecliptic plane orbit of Enceladus do not create a significant time-varying magnetic field in the moon's fixed frame. A recent analysis by \citeA{styczinski2024planetmag} estimated a harmonic external field at the orbital period. In a detailed analysis dedicated to Enceladus by \citeA{Saur2024}, several excitation mechanisms with different periods and their subsequent induction responses within a conductive interior were studied. They showed that external time-varying fields are present in the Cassini data, and that small-amplitude magnetic field perturbations within the large plasma magnetic field perturbations are consistent with induction from a highly conductive interior. Relatively small induction signals, concealed by ubiquitous strong plasma-induced magnetic fields \cite{Saur2024}, impose additional requirements on the science payload and make the induction sounding of Enceladus more challenging compared to the Jovian system. 

Thanks to the Cassini-Huygens mission, Enceladus is among the most well-studied icy moons \cite{glein2018geochemistry, choblet2022enceladus}. In addition to its subsurface ocean, two other aspects make it a particularly interesting target for EM induction and motivate our study:
\begin{itemize}
    \item First, a low-density core requires a significant porosity \cite{choblet2017powering}, while the putative hydrothermal activity \cite{glein2018geochemistry, postberg2011salt} indicates permeability and fluid content \cite{GueguenPalciauskas1994}. Conductivity is highly sensitive even to small amounts of interconnected pore fluids \cite{glover2010generalized}. Therefore, we quantify the effects of porosity and pore-fluid temperature on electrical conductivity and infer a plausible conductivity range, showing that conductivity is useful for constraining the thermal and compositional properties of the core. 
    \item Second, there is evidence for a heterogeneous nonaxisymmetric structure of Enceladus's ice shell with significantly thinner ice at the poles than at the equator \cite{Cadek2019, hemingway2019enceladus}. This poses a question: can 3-D EM induction effects resulting from a heterogeneous ice shell be detected and used to inform us about the properties of the hydrosphere? To address this question, conventional electromagnetic induction techniques that are mostly limited to 1-D models are insufficient. Therefore, we applied a new geophysical 3-D electromagnetic induction solver to model the actual shape and heterogeneous ice-shell thickness.
\end{itemize}
Therefore, the primary objective of this study is to quantify the global and local EM induction responses and fields for a range of plausible interior conductivity models at periods that can probe the conductivity structure of the ocean and core. 

We begin the study by developing an interior electrical conductivity model of Enceladus in Section \ref{sec:interior}. We model the ocean's conductivity as that of a saline fluid and develop a bulk conductivity model for the two-phase porous core. We then elaborate on the physical model of the electromagnetic induction in Section \ref{sec:induction}, starting with a 1-D interior scenario and then proceeding to the general case of a 3-D electrical conductivity distribution. We discuss the relevant EM transfer functions and briefly outline the finite-element EM modelling approach adopted for the 3-D case. We present the modeling results in Section \ref{sec:results} and conclude the study with a discussion of the findings in the context of current and future missions in Section \ref{sec:disc}.

\section{Materials and Methods}
\label{sec:methods}

\subsection{Interior model of Enceladus}
\label{sec:interior}

\begin{table}
\centering
\caption{Parameters of radial (1-D) and laterally heterogeneous (3-D) models.}
\begin{tabular}{|c|c|c|c|}
\hline
Model type & Surface radius & Ice thickness & Core radius \\ \hline
1-D        & 252 km         & 21 km         & 193 km      \\ \hline
3-D        & \citeA{Tajeddine2017} & \citeA{Cadek2019} &  193 km \\ \hline
\end{tabular}
\label{tab:params}
\end{table}

\begin{figure}
\noindent\includegraphics[width=\textwidth]{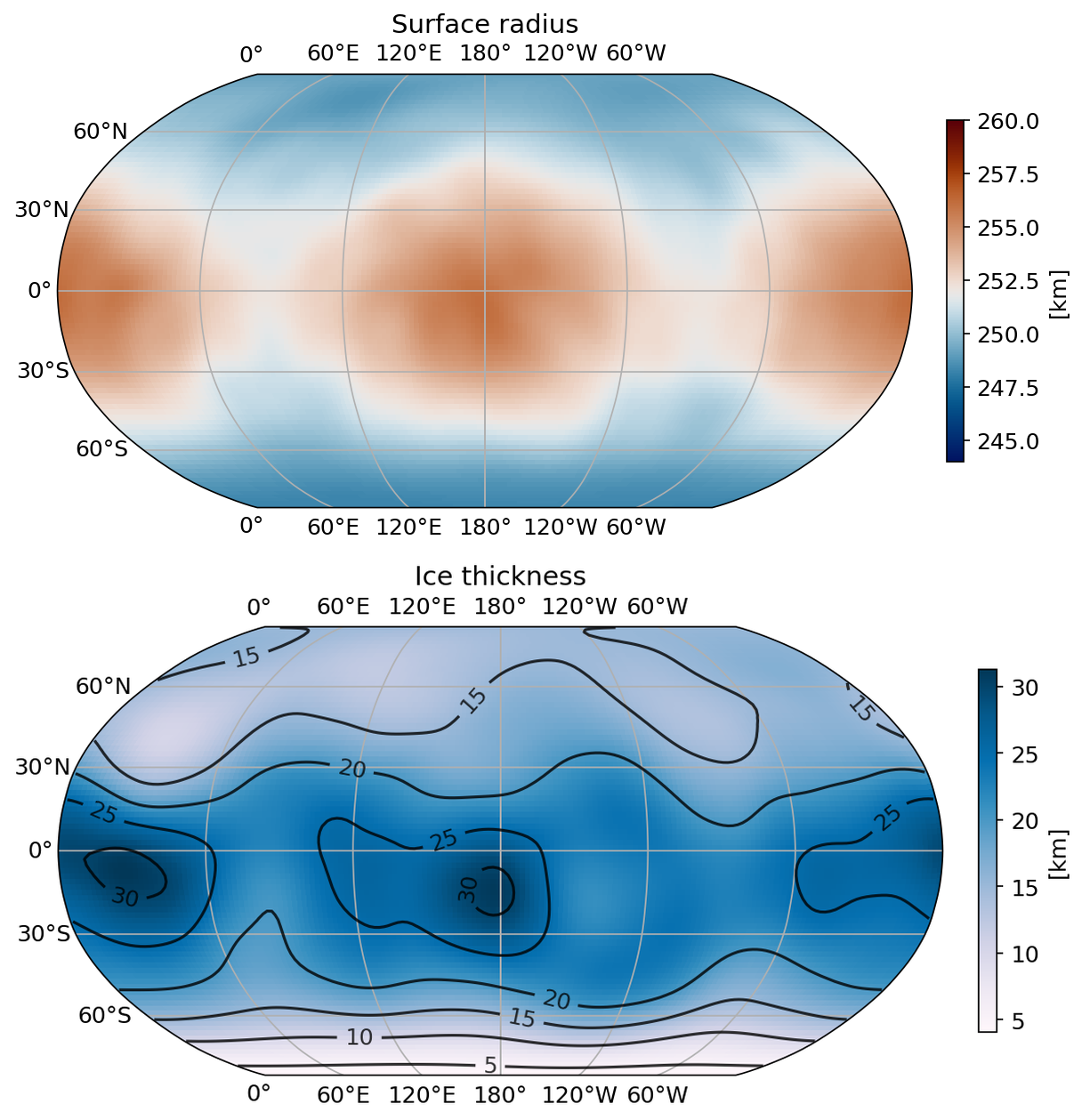}
\caption{Top: surface radius of Enceladus following the model of \citeA{Tajeddine2017}. Bottom: ice shell thickness model of \citeA{Cadek2019}, truncated to spherical harmonic degrees $l \leq 6$. The longitude is oriented from the west to the east, whereby the 180$^{\circ}$ meridian corresponds to the center of the anti-Saturnian hemisphere.}
\label{fig:surface_ice}
\end{figure}

We split Enceladus's interior into three regions: an ice shell overlying the ocean, and an underlying core (Figure \ref{fig:mesh}). In the 1-D scenario, we assume that the ice and ocean are spherical shells of a constant thickness as specified in Table \ref{tab:params}. In the 3-D scenario, the thickness of the ice shell (and hence the ocean) is inferred from the models of \citeA{Tajeddine2017, Cadek2019} depicted in Figure \ref{fig:surface_ice}. For both 1-D and 3-D models, the core has a constant radius of 193 km. Note that the model of \citeA{Cadek2019} predicts undulations of the core radius up to $\approx 2.5$ km around the mean value of 193 km. We found these differences negligible for electromagnetic induction responses and, therefore, used a constant core radius throughout. We used the body-fixed (planetocentric) frame for Enceladus, with $+Z$ pointing toward Enceladus’ north pole, $+X$ pointing to Saturn, and $+Y$ completing a right-handed set, i.e., $90^{\circ}$ east of $+X$ in the equatorial plane.

\begin{table}[h!]
\centering
\caption{Conductivity values $[\mathrm{Sm}^{-1}]$ of the porous rocky core ($\sigma_{core}$) and ocean ($\sigma_{ocean}$) layers used for the four respective model scenarios. Ice shell conductivity of $10^{-5}$ S/m is used throughout.}
\begin{tikzpicture}
\node (A) at (0,0) {
    \begin{tabular}{|c|c|c|}
    \hline
    \begin{tabular}{@{}c@{}}\rule{0pt}{1em}\hspace{2.5em}\\[-1.5em]
        \begin{tikzpicture}[baseline=(X.base)]
        \node[inner sep=1pt] (X) at (0,0) {\phantom{abc}};
        \draw[thick] (-1,0.6) -- (1,-0.6);
        \node[anchor=north east] at (1,0.5) {\textit{$\sigma_{core}$}};
        \node[anchor=south west] at (-1,-0.5) {\textit{$\sigma_{ocean}$}};
        \end{tikzpicture}\\[0.2em]
    \end{tabular}
    & \textbf{0.01} $[\mathrm{Sm}^{-1}]$ & \textbf{10} $[\mathrm{Sm}^{-1}]$ \\
    \hline
    \textbf{0.5} $[\mathrm{Sm}^{-1}]$ & {Model 1} & {Model 2} \\
    \hline
    \textbf{5} $[\mathrm{Sm}^{-1}]$ & {Model 3} & {Model 4} \\
    \hline
    \end{tabular}
};
\end{tikzpicture}
\label{tab:models}
\end{table}

We model electromagnetic induction responses for both 1-D and 3-D models using two different electrical conductivity values for ocean and core, resulting in four reference models listed in Table \ref{tab:models}. An ice electrical conductivity of $10^{-5}$ S/m was used for all models, a value reported for terrestrial polar ice \cite{barnes2002effect}. The range of conductivity values used for the ocean and the core in Table \ref{tab:models} encompasses a wide spectrum of geophysically plausible scenarios. With the temperature of Enceladus’s ocean between $-2^{\circ}$C and $1^{\circ}$C \cite{Zolotov2007, glein2018geochemistry}, its electrical conductivity is primarily determined by the amount and composition of dissolved salts and volatiles. The lower value of $0.5$ S/m that we used corresponds to the conductivity of NaCl-H$_2$O or seawater salt solutions with an absolute salinity of $\approx 10$ g/kg at relevant temperatures \cite{sen1992influence}. For higher salinity values \cite{Kang2022} and/or in the presence of additional conductivity-enhancing solubles \cite{CastilloRogez2022}, the ocean conductivity can reach values of up to 5-7 S/m. A recent study by \citeA{Glein2025} predicts a similar range of $0.83 - 7.5$ S/m for the ocean's conductivity based on geochemical modeling and plume observations. Following these studies, we set 5 S/m as the upper bound for ocean conductivity in our simulations.

\begin{figure}
\centering\includegraphics[width=0.7\textwidth]{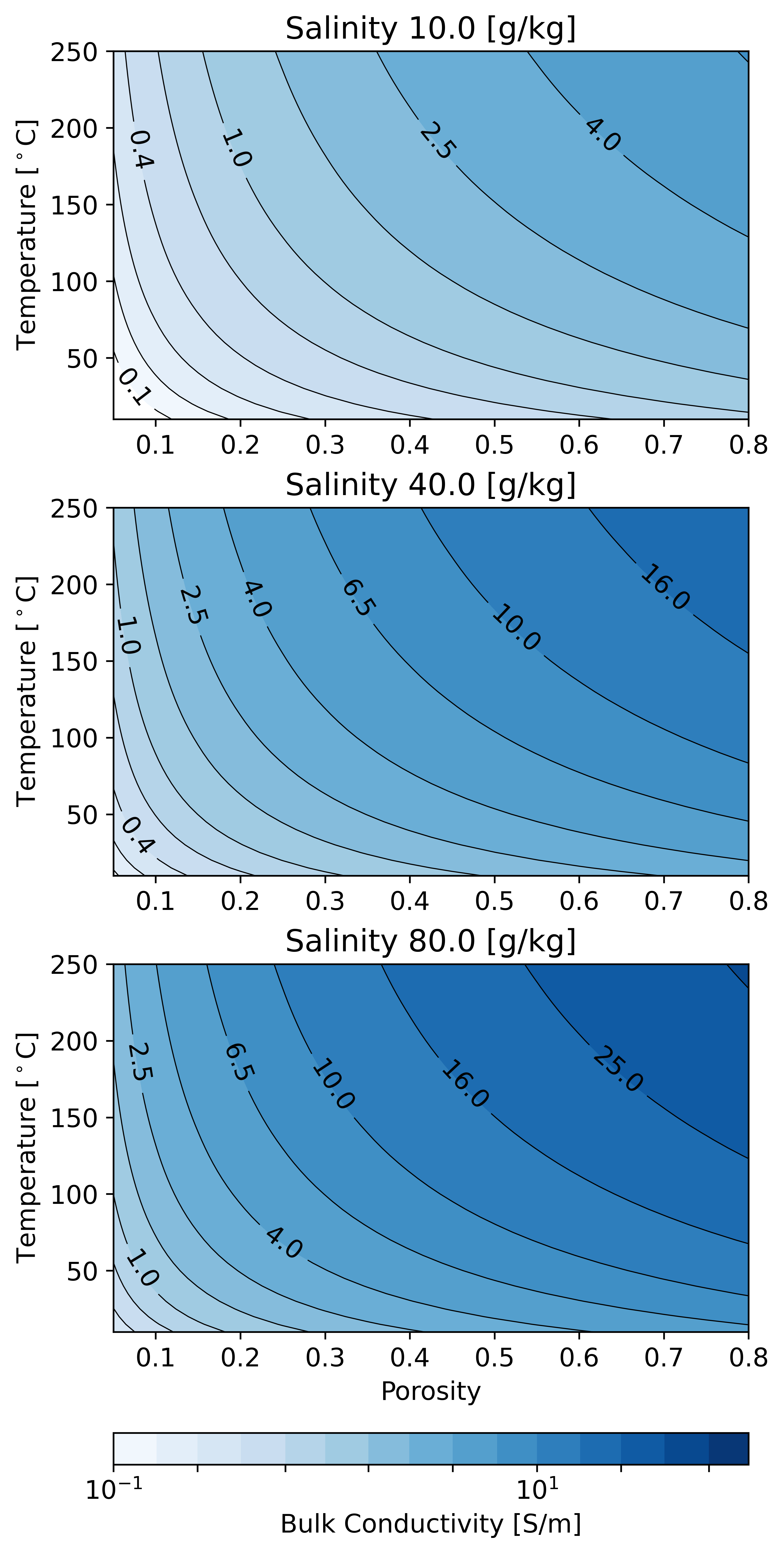}
\caption{Bulk electrical conductivity of a porous core as a function of porosity and temperature for selected salinity values. The bulk conductivity was computed using the Hashin-Shtrikman upper bound for a two-phase medium, assuming a well-interconnected pore space.}
\label{fig:bulk_conductivity}
\end{figure}

Since the core properties are poorly constrained, we chose a $0.01 - 10$ S/m range to represent the bulk conductivity of the core. The actual value will depend on the salinity and temperature of the pore fluid, as well as the porosity and interconnectedness of the porous space \cite{GueguenPalciauskas1994}, which are also related to permeability. Enceladus's core is believed to be porous \cite{choblet2017powering, kisvardai2023investigating} and hydrothermally active \cite{Bouffard2025, Sekine2015}. The latter implies that (i) pore fluids are at higher temperatures compared to the global ocean and (ii) the pores are interconnected to ensure the permeability levels that are sufficient to sustain the hydrothermal circulation \cite{spinelli2004hydrothermal}. For instance, \citeA{choblet2017powering} used homogeneous porosities up to 30\%, whereas \citeA{kisvardai2023investigating} allowed for variations in porosity with depth, resulting in higher porosities ($\approx 40$\%) close to the core-ocean boundary. Furthermore, temperatures reaching a few hundred degrees Celsius deep in the core have been previously discussed \cite{Hsu2015, Sekine2015, choblet2017powering, glein2018geochemistry}. Figure \ref{fig:bulk_conductivity} shows the bulk conductivity of the core as a function of temperature and porosity for different salinity values. The NaCl-H$_2$O solution for the fluid \cite{sen1992influence} and the Hashin-Shtrikman upper bound \cite{hashin1962variational} representing a two-phase medium with the interconnected pore space were used. A conductivity of $10^{-3}$ S/m is applied for the solid (rock) phase. The bulk electrical conductivity of $\approx 10$ S/m is reached at $\approx 200^{\circ}$C for a moderate salinity of 40 [g/kg] and high porosities of 50\%. However, these values are largely unconstrained by existing observations. Therefore, the conductivity limit of $0.01$ (respectively 10) S/m we used for the core bulk conductivity here corresponds to plausible scenarios of low (respectively high) salinity, temperature, and porosity. We note that an additional mechanism that can further enhance the conductivity of the core is hydrothermal alteration of minerals, leading to the formation of electrically conductive clays \cite{aoyama2024numerical}.

\subsection{Electromagnetic induction}
\label{sec:induction}

In and at the surface of Enceladus, we describe space and time variations of the electromagnetic field using Maxwell's equations with the displacement current neglected:
\begin{eqnarray}
\mu^{-1}\nabla \times \mathbf{\tilde B} &=& \sigma\mathbf{\tilde E} + \mathbf{\tilde J}^{\text{ext}}, \label{eq:maxwell_time1} \\
\nabla \times \mathbf{\tilde E} &=& -\frac{\partial\mathbf{\tilde B}}{\partial t}, \label{eq:maxwell_time2}
\end{eqnarray}
where $\mu \equiv \mu_0 = 4\pi \times 10^{-7}$ [H/m] is the magnetic permeability of vacuum, $\sigma$ [S/m] electrical conductivity; $\mathbf{\tilde B}$ [T], $\mathbf{\tilde E}$ [V/m] are magnetic and electric fields, respectively. Position in a body-fixed reference frame is given by the vector $\vec{r}=(r, \theta, \varphi)$ where $r$, $\theta$, and $\varphi$ are the radial distance, co-latitude, and longitude, respectively. $\mathbf{\tilde J}^{ext}$ [A/m$^2$] is the extraneous source current density.

By assuming the following Fourier convention
\begin{equation}
\tilde{F}(t)=\frac{1}{2\pi}\int\limits_{-\infty}^{\infty}F(\omega)e^{-\mathrm{i}\omega t}\mathrm{d}\omega
\label{eq:fourier}
\end{equation}
where $\omega$ denotes the angular frequency, the governing equations can be written in the frequency domain as
\begin{eqnarray}
\mu^{-1}\nabla \times \mathbf{B} &=& \sigma\mathbf{E} + \mathbf{J}^{\text{ext}}, \label{eq:maxwell_fd1} \\
\nabla \times \mathbf{E} &=& \mathrm{i}\omega\mathbf{B}, \label{eq:maxwell_fd2}
\end{eqnarray}

Outside Enceladus and in regions with $\mathbf{J}^{\text{ext}} = 0$, the classic potential field representation \cite{parkinson1983} of the magnetic field yields
\begin{eqnarray}
\label{eq:magpotential}
  \mathbf{\tilde B}(\vec{r},t; \sigma) = -\nabla \left(\tilde{V}^{\text{ext}}(\vec{r},t) + \tilde{V}^{\text{int}}(\vec{r},t;\sigma)\right),
\end{eqnarray}
where scalar potentials due to external and internal (induced) sources are
\begin{eqnarray}
\label{eq:magpotentials_ext}
  \tilde{V}^{\text{ext}}(\vec{r}, t) &=& \textnormal{Re}\left\{R \sum_{n=1}^{\infty}\sum_{m = -n}^{n} \tilde{\varepsilon}_n^m(t) \left( \frac{r}{R} \right)^n Y_n^m(\theta, \varphi)\right\} \\
  &=& R\sum_{n = 1}^{\infty}\sum_{m = 0}^n \left[ \tilde{q}_n^m(t)\cos(m\varphi) + \tilde{s}_n^m(t)\sin(m\varphi) \right]\left( \frac{r}{R} \right)^n P_n^m(\cos{\theta}),
\end{eqnarray}
and
\begin{eqnarray}
\label{eq:magpotentials_int}
  \tilde{V}^{\text{int}}(\vec{r}, t; \sigma) &=& \textnormal{Re}\left\{R \sum_{k=1}^{\infty}\sum_{l = -k}^{k} \tilde{\iota}_k^l(t; \sigma) \left( \frac{r}{R} \right)^{-(k+1)} Y_k^l(\theta, \varphi)\right\} \\
  &=& R\sum_{k = 1}^{\infty}\sum_{l = 0}^k \left[ \tilde{g}_k^l(t;\sigma)\cos(l\varphi) + \tilde{h}_k^l(t; \sigma)\sin(l\varphi) \right]\left( \frac{r}{R} \right)^{-(k+1)} P_k^l(\cos{\theta}),
\end{eqnarray}
respectively. Here, $Y_n^m(\theta, \phi) = P_n^{|m|}(\cos{\theta})\exp{(\textnormal{i}m\varphi)}$ are the complex Spherical Harmonic functions of degree $n$ and order $m$ with $P_n^{|m|}(\cos{\theta})$ being the Schmidt semi-normalized associated Legendre function \cite{winch2005geomagnetism}, $R$ is the reference sphere radius that encloses the region of internal sources and ensures correct units for $\mathbf{\tilde B}$. The $\tilde{\varepsilon}_n^m, \tilde{\iota}_n^m$ are external and internal complex Gauss coefficients, respectively. They are related to real-valued Gauss coefficients as
\begin{equation}
    \tilde{q}_{n}^0 = \varepsilon_n^0,\qquad 
    \left\{\begin{aligned}
        &\tilde{q}_n^m = 2\mathrm{Re}[\tilde{\varepsilon}_n^m] \\ 
        &\tilde{s}_n^m = -2\mathrm{Im}[\tilde{\varepsilon}_n^m],
    \end{aligned}\right. \quad m > 0.
\end{equation}
and similarly for the internal coefficients $\tilde{\iota}_k^l$ and $\tilde{g}_k^l,\tilde{h}_k^l$. Note that the different indexing between external and internal coefficients is introduced to accommodate the general case with a 3-D electrical conductivity, which we discuss in subsequent sections.

We note that the physical model developed here assumes no galvanic coupling between the interior and exterior of Enceladus. Even if some ionized particles exist near the surface and the ice shell may contain local cavities or fractures filled with a conductive material, electromagnetic induction remains the dominant global mechanism driving electric currents in the interior, since only a negligible electric current can flow directly through a highly resistive ice shell \cite{saur2010induced}.

\subsubsection{Radially symmetric 1-D case}

Assuming that the conductivity changes only with the radial distance, that is $\sigma \equiv \sigma(r)$, and the external field is described by the coefficient $\varepsilon_n^m(\omega)$, we can write the total magnetic and electric field at the surface as \cite{kuvshinov2012global}
\begin{eqnarray}
_{n}^{m}B_r(\vec{r}_{R^-}, \omega; \sigma)&=& \varepsilon_n^m(\omega)(2n+1)\frac{nZ_n}{i\omega\mu_0 R - n Z_n} Y^m_n(\theta,\varphi),\label{eq:br_impedance}\\
_{n}^{m}\mathbf{B}_{H}(\vec{r}_{R^-}, \omega; \sigma) &=&-\varepsilon_n^m(\omega)\frac{2n+1}{n+1} \frac{i\omega\mu_0 R}{i\omega\mu_0 R - n Z_n } \nabla_{\perp} Y^m_n(\theta,\varphi),\label{eq:btau_impedance} \\
_n^m\mathbf{E}_{H}(\vec{r}_{R^-}, \omega; \sigma) &=&\varepsilon_n^m(\omega)\frac{1}{\mu_0}
\frac{2n+1}{n+1} \frac{i\omega\mu_0 R Z_n}{i\omega\mu_0 R - n Z_n}\hat{e}_r\times \nabla_{\perp} Y^m_n(\theta,\varphi),\label{eq:e_impedance}
\end{eqnarray} 
where $Z_n \equiv Z_n({R^-}, \omega; \sigma)$ is the spectral impedance of a spherical conductor \cite{srivastava1966theory} and $\vec{r}_R = (r\rightarrow R^-, \theta, \phi)$ denotes a position at the surface of a sphere with radius $R = 252$ km where $r$ approaches the surface from below. The subscript $_H$ denotes the horizontal component. The surface gradient is given by
\begin{equation}
\label{eq:nabla_perp}
\nabla_{\perp} = \hat{e}_{\theta}\frac{\partial}{\partial\theta} + \hat{e}_{\phi}\frac{1}{\sin{\theta}}\frac{\partial }{\partial\varphi}.
\end{equation}
Note that if the external time-varying field is described by more than one SH coefficient, the total field is a mere superposition of individual SH modes given by eqs. (\ref{eq:br_impedance}-\ref{eq:e_impedance}).

The spectral impedance is related to the so-called $C$-response as 
\begin{equation}
\label{eq:C2Z}
Z_n = -\mathrm{i}\omega\mu C_n.
\end{equation}
$C$-response is a commonly used EM transfer function in geomagnetism. It has units of meters, its real part is a monotonic function of period for a radially symmetric model, and represents the "center-of-mass" of the induced current density in the interior \cite{weidelt1972inverse}. Furthermore, using eqs. (\ref{eq:br_impedance}-\ref{eq:C2Z}) and assuming that $Y_n^m$ describes the magnetic field geometry, allows us to derive the $C$-response at a location $\vec{r}_R$ on the surface via horizontal and radial magnetic field components as
\begin{align}
C_n(R, \omega; \sigma) &= \frac{1}{\sin{\theta}}\frac{R}{n(n+1)}\frac{\partial_{\varphi}Y_n^m(\theta, \varphi)}{Y_n^m(\theta, \varphi)}\frac{{B_{r}(\vec{r}_R,\omega; \sigma)}}{{B_{\phi}(\vec{r}_R,\omega; \sigma)}} \label{eq:c_via_h1} \\ 
&= \frac{R}{n(n+1)}\frac{\partial_{\theta}Y_n^m(\theta, \varphi)}{Y_n^m(\theta, \varphi)}\frac{B_r(\vec{r}_R,\omega; \sigma)}{B_{\theta}(\vec{r}_R,\omega; \sigma)}. \label{eq:c_via_h2}
\end{align}
Note that the $C$-response depends on the radius, frequency, and radial conductivity distribution, as well as on the SH degree of the external (inducing) magnetic field. Using eqs. (\ref{eq:br_impedance}-\ref{eq:btau_impedance}) and (\ref{eq:C2Z}) with eqs. (\ref{eq:c_via_h1}-\ref{eq:c_via_h2}), one can show that the C-response is constant everywhere on the surface of a radially symmetric conductor. The eqs. (\ref{eq:c_via_h1}-\ref{eq:c_via_h2}) are particularly useful, as they allow the estimation of EM transfer functions using local magnetic field measurements. Therefore, $C$-response forms the basis of many electromagnetic sounding studies of the Earth \cite{grayver2024unravelling}. 

Alternatively to using magnetic field components, the local $C$-response can be expressed from the ratio of horizontal electric and magnetic field components as
\begin{align}
C_n(R, \omega; \sigma) &= -\frac{1}{\mathrm{i}\omega}\frac{{E_{\theta}(\vec{r}_R,\omega; \sigma)}}{{B_{\phi}(\vec{r}_R,\omega; \sigma)}} \nonumber \\ 
&= \frac{1}{\mathrm{i}\omega}\frac{E_{\phi}(\vec{r}_R,\omega; \sigma)}{B_{\theta}(\vec{r}_R,\omega; \sigma)}. \label{eq:c_via_e}
\end{align}
In theory, transfer functions obtained using magnetic-to-magnetic or electric-to-magnetic ratios are identical. However, as we will demonstrate below, there are significant practical implications depending on the approach adopted.

Next, we can introduce another transfer function that is related to the $C$-response and represents a spectral ratio of internal and external SH coefficients \cite{schmucker1985magnetic, parkinson1983}
\begin{eqnarray}
\label{eq:Qresponse}
Q_n(\omega;\sigma) &=& \frac{n}{n+1}\frac{R - (n+1)C_n(R, \omega; \sigma)}{R + n C_n(R, \omega; \sigma)} \nonumber \\ 
&=& \frac{\iota_n^m(\omega; \sigma)}{\varepsilon_n^m(\omega)}.
\end{eqnarray}
Note that $Q_n$ is independent of location and order $m$. Further, it shows that in the 1-D case, each external coefficient induces only one internal coefficient of the same degree and order. In the time domain, the $Q_n$ transfer function represents an impulse response, such that internal and external coefficients can be related through a convolution integral \cite{schmucker1985magnetic}
\begin{eqnarray}
\label{eq:Qresponse_td}
\tilde{\iota}_n^m(t; \sigma) &=& \int_{-\infty}^t \tilde{Q}_n(t - \tau; \sigma)  \tilde{\varepsilon}_n^m(\tau)\textnormal{d}\tau.
\end{eqnarray}
This equation shows that for a body with finite conductivity, the induced coefficient at time $t$ depends not only on the instantaneous external (inducing) field coefficient, but also on the external coefficient at earlier times. The decay rate of the impulse response $\tilde{Q}_n$ is determined by the subsurface conductivity profile \cite{grayver2021time}. Combining eqs. (\ref{eq:magpotential}-\ref{eq:magpotentials_int}) and (\ref{eq:Qresponse}), we can express the magnetic field due to the coefficient $\varepsilon_n^m(\omega)$ outside the body ($r \geq R$) as \cite{grayver2021time}
\begin{eqnarray}
\label{eq:BviaQ}
_{n}^{m}B_{r}(\vec{r}_{r \geq R},\omega;\sigma) &=& -\varepsilon_{n}^{m}(\omega)
\Bigl[ n\left(\frac{r}{R}\right)^{n-1} - (n+1)\left(\frac{R}{r}\right)^{n+2}\,Q_{n}(\omega;\sigma)\Bigr]Y_{n}^{m}(\theta,\varphi), \nonumber \\
_{n}^{m}\mathbf{B}_H(\vec{r}_{r \geq R},\omega;\sigma) &=& -\varepsilon_{n}^{m}(\omega)
\Bigl[ \left(\frac{r}{R}\right)^{n-1} + \left(\frac{R}{r}\right)^{n+2}Q_{n}(\omega;\sigma) \Bigr] \nabla_{\perp} Y_{n}^{m}(\theta,\varphi).
\end{eqnarray}

The formalism developed above fully elaborates the electromagnetic response of a radially symmetric body. It establishes the relationship between the transfer functions, the magnetic (electric) field on the surface, and the potential representation outside the conductor. More importantly, it equips us with a very powerful toolset for sounding icy moon interiors using EM field observations. Eqs. (\ref{eq:br_impedance}-\ref{eq:btau_impedance}) and (\ref{eq:c_via_h1}-\ref{eq:c_via_e}) allow us to estimate the $C_n$ transfer function from the magnetic and, if available, electric fields at the surface. These equations are most useful when local, continuous observations of EM fields are available, as would be the case with a lander. In contrast, Eqs. (\ref{eq:BviaQ}) are useful for estimating the $Q_n$ transfer function using magnetic field observations from a satellite orbiter \cite{Olsen1999}. For a 1-D body, the $C_n$ and $Q_n$ response functions are related via eq. (\ref{eq:Qresponse}). 

Because of its convenience in the context of space-based observations, the $Q$-response has been the most commonly used transfer function in the planetary and icy moons community, often expressed in terms of its amplitude $A_n$ and phase $\Phi_n$ as
\begin{equation}
\label{eq:Qampphase}
    Q_n = n/(n+1)A_n\exp(\mathrm{i}\Phi_n).
\end{equation}

\begin{figure}
\noindent\includegraphics[width=\textwidth]{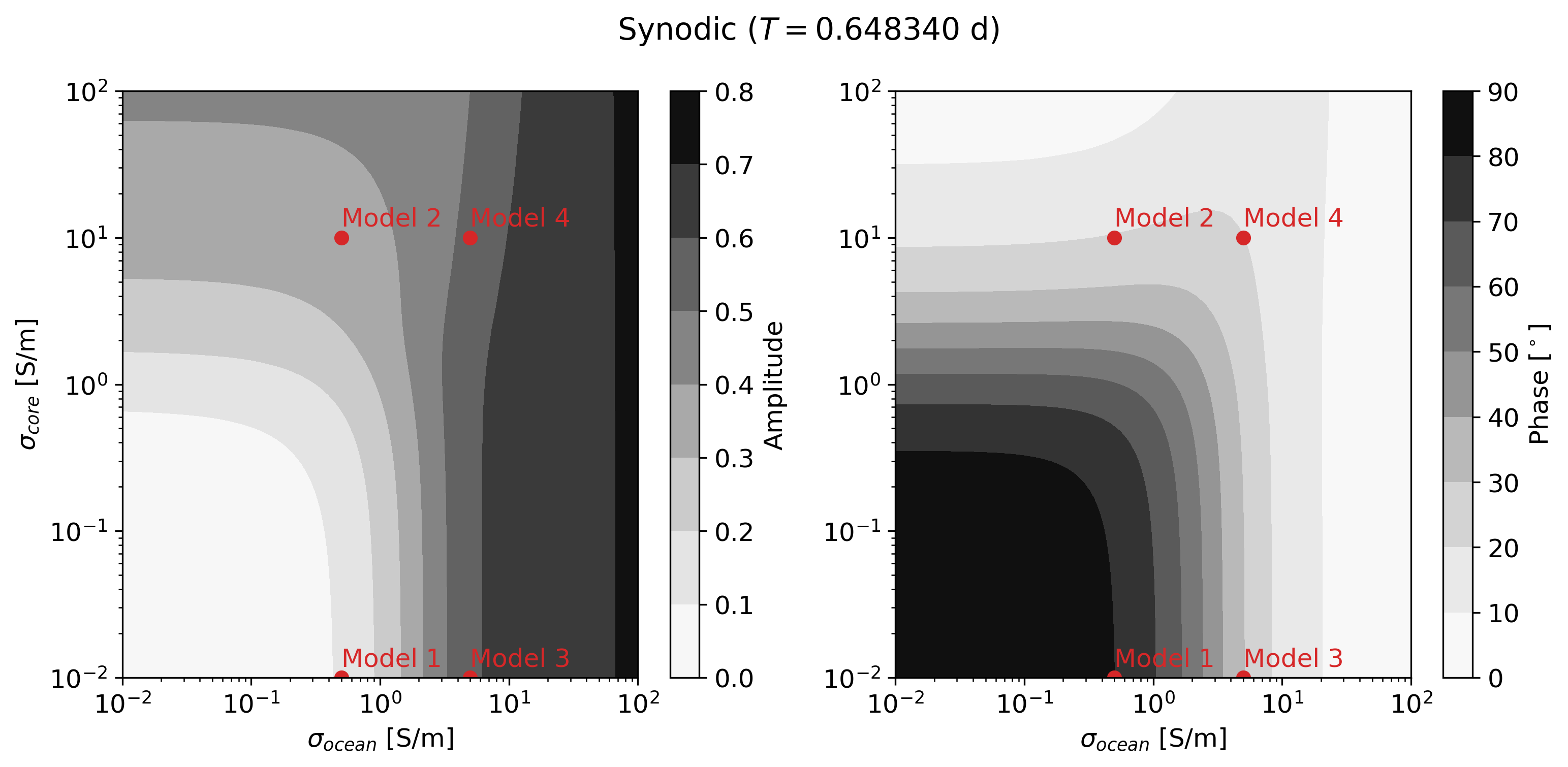}
\caption{Amplitude and phase of the degree one ($n=1$) induction transfer function at the synodic period for a three-layer radially symmetric (1-D) model of Enceladus (for model definition, refer to Table \ref{tab:params}), calculated as a function of the core and ocean layer conductivities. Circles depict four reference conductivity models from Table \ref{tab:models}.}
\label{fig:q1d}
\end{figure}

Before proceeding to a general 3-D case, we note that at frequencies where the spatial variations of the external field and curvature effects are small compared to the EM field diffusion length, the spherical impedance $Z_n$ will approach the plane-wave impedance of a layered medium \cite{srivastava1966theory, berdichevsky2002magnetotellurics}. The diffusion length is typically expressed by skin depth, but a more accurate proxy is $|C_n|$, which designates the approximate depth of the maximum induced current in a radially symmetric conductor \cite{weidelt1972inverse}. Then, the asymptotic plane-wave approximation at the surface is justified if
\begin{equation}
\label{eq:planewave_cond}
    |C_n| \ll R/n.
\end{equation}
In this work, we do not make the plane-wave assumption and model external and induced fields in spherical geometry. However, a more detailed discussion on the plane-wave approximation and its implications for EM induction sounding is given in \ref{app:planewave}.

\subsubsection{General 3-D case}

The formalism presented in the previous section is valid under the assumption that the subsurface conductivity varies only with the radial distance. Proceeding to the general case with $\sigma \equiv \sigma(\vec{r})$ invalidates many of the equations in the previous section, at least in the sense of being an equation and not an approximation. Specifically, the electromagnetic response of a laterally heterogeneous object to a harmonic time-varying external magnetic field can no longer be expressed through a single scalar transfer function such as $Q_n$ or spectral impedance $Z_n$. In a 3-D object, each external coefficient $\varepsilon_n^m$ induces a series of internal coefficients $\iota_k^l$ which are related through a so-called $Q$-matrix \cite{Olsen1999} as
\begin{equation}
\label{eq:qmatrix}
\iota_k^l(\omega; \sigma) = \sum_{n=1}^{\infty}\sum_{m=-n}^{n}Q_{kn}^{lm}(\omega; \sigma)\varepsilon_n^m(\omega).
\end{equation}
In case of a 1-D conductivity, the $Q$-matrix becomes a diagonal matrix with $Q_n$ elements from eq. (\ref{eq:Qresponse}) on the main diagonal. In Earth studies, the $Q$-matrix has been used to probe lateral conductivity variations within the Earth's mantle \cite{kuvshinov2021probing} and to model 3-D induction effects due to bathymetry and heterogeneity in the ocean and sediments \cite{grayver2021time}. 

Therefore, the $Q$-matrix represents the full electromagnetic response of a body with an arbitrary 3-D conductivity distribution to an external harmonic magnetic field described by $\varepsilon_n^m(\omega)$. We can use eqs. (\ref{eq:magpotentials_ext}-\ref{eq:magpotentials_int}) together with eq. (\ref{eq:qmatrix}) to write the total magnetic field in frequency domain as \cite{puthe2014mapping}
\begin{eqnarray}
\label{eq:BviaQmatrix}
_{n}^{m}B_{r}(\vec{r}_{r \geq R},\omega;\sigma) &=& -\varepsilon_{n}^{m}(\omega)
\Bigl[ n\left(\frac{r}{R}\right)^{n-1}Y_{n}^{m}(\theta,\varphi) -\sum_{k = 1}^{\infty}\sum_{l = -k}^{k}(k+1) \left(\frac{R}{r}\right)^{k+2}Q_{kn}^{lm}(\omega;\sigma)Y_{k}^{l}(\theta,\varphi)\Bigr], \nonumber \\
_{n}^{m}\mathbf{B}_H(\vec{r}_{r \geq R},\omega;\sigma) &=& -\varepsilon_{n}^{m}(\omega)
\Bigl[ \left(\frac{r}{R}\right)^{n-1}\nabla_{\perp} Y_{n}^{m}(\theta,\varphi) + \sum_{k = 1}^{\infty}\sum_{l = -k}^{k}\left(\frac{R}{r}\right)^{k+2}Q_{kn}^{lm}(\omega;\sigma)\nabla_{\perp} Y_{k}^{l}(\theta,\varphi) \Bigr].
\end{eqnarray}
The equivalent time-domain representation is rather cumbersome and lengthy, but has been previously elaborated by \citeA{grayver2021time}. Note that the reference radius $R$ is not necessarily the same between the 1-D and 3-D cases. For the 1-D model, we used $R = 252$ km, corresponding to the mean radius of Enceladus. However, our 3-D model includes surface undulations (Figure \ref{fig:surface_ice}) that extend to the radial distance of $\approx 256.2$ km, representing the radius of the minimal bounding sphere -- that is, the minimum radius of a spherical surface on which the potential field representation from eq. (\ref{eq:magpotential}) is permissible.

Furthermore, the $C$-response calculated from the ratio of the local magnetic (electric) field components (eqs. \ref{eq:c_via_h1}-\ref{eq:c_via_e}) will no longer be constant everywhere on the surface of a non-axisymmetric model. In addition, its real part is no longer a monotonic function of period. Therefore, the lateral variability in the $C$-response over the surface is a proxy for lateral variations in the subsurface conductivity. On Earth, this property has allowed us to constrain the lateral variability of the mantle electrical conductivity using a network of ground observatories \cite{kelbert2009global, kuvshinov2012global}. 

\subsubsection{Numerical solution for 3-D case}

To compute the electromagnetic response of a heterogeneous 3-D object, we adapted the finite-element electromagnetic solver GoFEM \cite{grayver2019three} that was previously used to model 3-D EM induction in the solid Earth and oceans \cite{grayver2021time, grayver2024magnetic}, and to image the subsurface 3-D conductivity using large-scale EM array data \cite{munch2026multi}. The GoFEM adopts the so-called primary-secondary field formulation \cite{monk2003finite}, whereby the total EM field is represented as the sum 
\begin{eqnarray}
\label{eq:decomp}
    \mathbf{E} = \mathbf{E}^p + \mathbf{E}^s, \nonumber \\
    \mathbf{B} = \mathbf{B}^p + \mathbf{B}^s,
\end{eqnarray}
with primary fields given by (cf. \ref{eq:maxwell_fd1}-\ref{eq:maxwell_fd2})
\begin{eqnarray}
\mu^{-1}\nabla \times \mathbf{B}^p &=& \sigma^p\mathbf{E}^p + \mathbf{J}^{\text{ext}}, \label{eq:maxwellp_fd1} \\
\nabla \times \mathbf{E}^p &=& \mathrm{i}\omega\mathbf{B}^p, \label{eq:maxwellp_fd2}
\end{eqnarray}
where $\sigma^p \equiv \sigma^p(r)$ is a radially symmetric part of the conductivity model. The extraneous current density term $\mathbf{J}^{\text{ext}}$ is formulated such that it produces the external magnetic field of desired geometry. Specifically, we introduce the current flowing in a thin sheet embedded in an insulator at an arbitrary radius $b$ above the surface
\begin{equation}
_n^m\mathbf{J}^{\text{ext}} = \frac{\delta(r - b)}{\mu}\varepsilon_n^m\frac{2n + 1}{n + 1}\left(\frac{b}{R}\right)^{n-1}\mathbf{e}_r \times \nabla_{\perp}Y_n^m(\theta,\varphi),
\end{equation}
which produces the external magnetic field due to the corresponding SH coefficient $\varepsilon_n^m$ in the region below the sheet \cite{schmucker1985magnetic}. Here, $\delta$ is the Dirac delta function. We compute the $\mathbf{E}^p, \mathbf{B}^p$ analytically by using the recursion formulae of \citeA{kuvshinov2012global}.

\begin{figure}
\noindent\includegraphics[width=\textwidth]{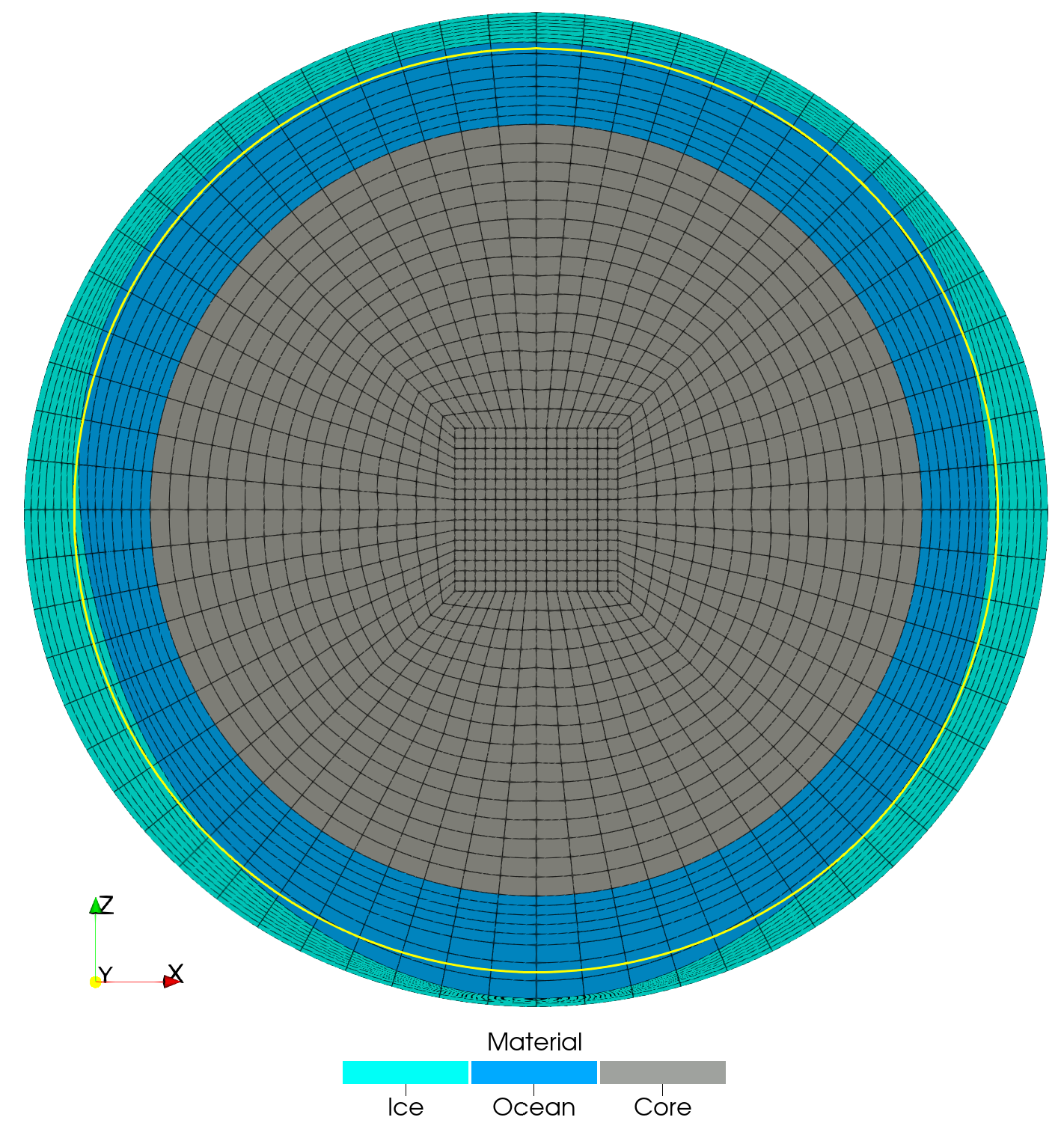}
\caption{Discrete mesh for the 3-D model with a heterogeneous ice shell thickness. The ice shell surface and base are constrained by models depicted in Figure \ref{fig:surface_ice} and described in Table \ref{tab:params}. The yellow circle depicts the ice base of the mean ice shell, corresponding to the 1-D model. Three colors denote ice, ocean, and core regions. Each region is assigned a homogeneous electrical conductivity according to Table \ref{tab:models}. The total number of cells in the mesh is 126976.}
\label{fig:mesh}
\end{figure}

Subtracting eqs. (\ref{eq:maxwellp_fd1}-\ref{eq:maxwellp_fd2}) from eqs. (\ref{eq:maxwell_fd1}-\ref{eq:maxwell_fd2}) yields
\begin{eqnarray}
\mu^{-1}\nabla \times \mathbf{B}^s &=& \sigma\mathbf{E}^s + \Delta\sigma\mathbf{E}^p, \label{eq:maxwells_fd1} \\
\nabla \times \mathbf{E}^s &=& \mathrm{i}\omega\mathbf{B}^s, \label{eq:maxwells_fd2}
\end{eqnarray}
where $\Delta\sigma = \sigma - \sigma^p$ is the laterally heterogeneous (3-D) part of the conductivity model. Taking the curl of eq. (\ref{eq:maxwells_fd2}), substituting eq. (\ref{eq:maxwells_fd1}) and rearranging results in the second-order PDE in terms of secondary (anomalous) electric field yields
\begin{equation}
\nabla \times \nabla \times \mathbf{E}^s - \mathrm{i}\omega\mu\sigma\mathbf{E}^s = \mathrm{i}\omega\mu\Delta\sigma\mathbf{E}^p \quad \textnormal{in}\quad  \Omega, \label{eq:esec}
\end{equation}
where $\Omega$ constitutes the modeling domain.
Since the surface of the model is not a perfect sphere and conductivity varies in all three dimensions, no local analytic boundary condition can be assigned at the boundary of the conductive domain $\Omega_{\sigma > 0}$ \cite{monk2003finite}. Therefore, we add an exterior shell with a negligible conductivity of $10^{-10}$ S/m, extending to a distance of $10R_E$, where $R_E$ is the mean radius of Enceladus. Assuming that the external boundary is sufficiently far, such that the tangential components of $\mathbf{E}^s$ approach zero, allows us to prescribe the homogeneous Dirichlet boundary conditions at the outer boundary
\begin{equation}
\mathbf{n} \times \mathbf{E}^s = 0 \quad \textnormal{on}\quad  \partial\Omega, \label{eq:bcs}
\end{equation}
where $\mathbf{n}$ is the unit normal to the surface $\partial\Omega$.
Eqs. (\ref{eq:esec}-\ref{eq:bcs}) form a Boundary-Value Problem (BVP) with the unique solution \cite{monk2003finite}. We solve the BVP using the finite-element method discretized on a hexahedral mesh (Figure \ref{fig:mesh}) that conforms to the undulated surfaces of the 3-D model (Table \ref{tab:models}). The corresponding secondary (anomalous) magnetic field $\mathbf{B}^s$ is computed from $\mathbf{E}^s$ via eq. (\ref{eq:maxwells_fd2}). The total field can then be obtained anywhere in $\Omega$ via (\ref{eq:decomp}). For more details about the finite-element solver, the reader is referred to \citeA{grayver2015large, grayver2019three}.

For this study, the radially symmetric part, $\sigma^p$, of the conductivity model consists of the core (given by a sphere of radius $193$ km) and the shell of $10^{-10}$ S/m that extends to the distance of $10R_E$. Accordingly, the 3-D component of the conductivity model $\Delta\sigma$, and hence the right-hand side of eq. \ref{eq:esec} are non-zero only within the ocean and ice regions. The advantage of the primary-secondary decomposition adopted here is twofold: (i) the primary part of the EM field is calculated analytically to a very high accuracy and only the secondary (anomalous) part is computed numerically, and (ii) since the right-hand side of eq. (\ref{eq:esec}) is not zero only in a confined volume within the domain $\Omega$ where $\Delta\sigma\neq 0$, we can use simple homogeneous boundary conditions (\ref{eq:bcs}) at the exterior boundary.

Notably, the literature on electromagnetic induction in icy moons often invokes the magnetic diffusion equation rather than the electric field formulation above. For completeness, we derive the magnetic diffusion equation to illustrate the equivalence. Dividing eq. (\ref{eq:maxwells_fd1}) by $\sigma$, taking the curl and substituting (\ref{eq:maxwells_fd2}) yields the magnetic diffusion equation in terms of the secondary magnetic field
\begin{equation}
\nabla \times \left(\eta\nabla \times \mathbf{B}^s\right) - \mathrm{i}\omega\mathbf{B}^s = \nabla \times \left(\sigma^{-1}\Delta\sigma\mathbf{E}^p\right) \quad \textnormal{in}\quad  \Omega, \label{eq:bsec}
\end{equation}
where $\eta = \frac{1}{\mu\sigma}$ is the magnetic diffusion coefficient. By analogy to the E-field formulation, solving eq. (\ref{eq:bsec}) and applying eq. (\ref{eq:maxwells_fd1}) yields $\mathbf{E}^s$ and $\mathbf{B}^s$ that can be used to obtain the total electric and magnetic field via eq. (\ref{eq:decomp}). Hence, the two formulations are equivalent. We note that outside Enceladus, the total magnetic field $\mathbf{B}$ from eq. (\ref{eq:decomp}) satisfies eq. (\ref{eq:BviaQmatrix}). The primary part of the field, $\mathbf{B}^p$, can be parameterized via eq. (\ref{eq:BviaQ}).

\section{Results}
\label{sec:results}

\citeA{Saur2024} and \citeA{styczinski2024planetmag} independently investigated several potential mechanisms of the external field variability around Enceladus at synodic ($T = 0.648340$ d) and orbital ($T = 1.370218$ d) periods. \citeA{Saur2024} found a uniform (that is, constant across the volume of Enceladus) time-varying external field to be consistent with Cassini's observations, albeit with remaining uncertainty on the external field polarization. Here, we write the external uniform time-harmonic magnetic field as a phasor
\begin{equation}
\mathbf{\tilde B}^{\mathbf{ext}}(t)=\Re\{\mathbf B^{\mathbf{ext}}\,e^{-i\omega t}\},\qquad
\mathbf{B}^{\mathbf{ext}}=\left(B_x(\omega),B_y(\omega),B_z(\omega)\right)
=\big(B_x^{0}e^{i\phi_x},\,B_y^{0}e^{i\phi_y},\,B_z^{0}e^{i\phi_z}\big),
\label{eq:extfield}
\end{equation}
which can be mapped to the SH basis as
\begin{equation}
q_1^0=-B_z,\qquad
q_1^1=-B_x,\qquad
s_1^1=-B_y.
\label{eq:mapping_phasor}
\end{equation}

\begin{figure}
\noindent\includegraphics[width=\textwidth]{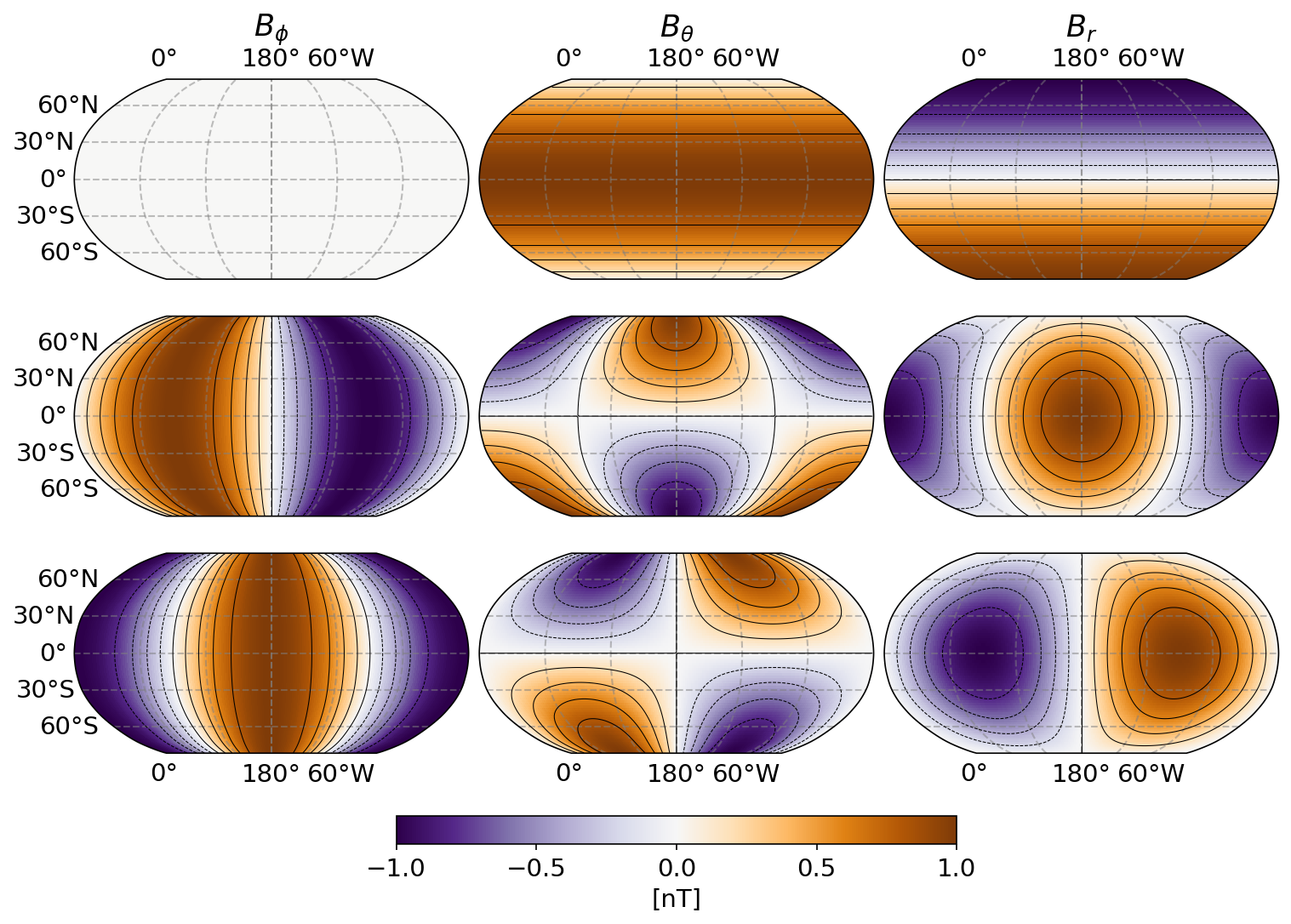}
\caption{Magnetic field components of the unit homogeneous external inducing field. Fields at the surface of a sphere with the reference radius are plotted. Each row corresponds to a field described (from top to bottom) by SH coefficients $q_1^0 = -B_z, q_1^1 = -B_x$, and $s_1^1 = -B_y$.}
\label{fig:Bext_dipole}
\end{figure}

In what follows, we model EM induction for three unit in-phase inducing field geometries, i.e., fields aligned with the three principal axes:
\begin{equation*}
B_{x,y,z}^0 = 1 \; \mathrm{nT}, \quad \phi_{x,y,z} = 0^{\circ}.
\end{equation*}
Because Maxwell's equations are linear with respect to the magnitude of the inducing external field, the computed unit responses can be scaled by the actual $\mathbf{B}^{\mathbf{ext}}$ to get in-situ responses with the actual magnitude.
Figure \ref{fig:Bext_dipole} shows the field components of these polarizations. Having EM induction due to these three fundamental inducing modes allows us to represent any uniform external field by invoking eqs. (\ref{eq:extfield}-\ref{eq:mapping_phasor}) and applying actual amplitudes and phases. For brevity, we show the results for $q_1^0$ in the main text, and for $q_1^1$ in the SI. The results for $s_1^1$ are very similar to $q_1^1$ (subject to the 90$^{\circ}$ rotation) and hence are omitted.

\subsection{Global fields}
\label{sec:global}

Before discussing the 3-D simulations, we first analyze the $Q_1$ transfer function computed for the three-layer 1-D model (Table \ref{tab:params}) and shown in Figure \ref{fig:q1d} as amplitude and phase (eq. \ref{eq:Qampphase}) at synodic period. For the low-conductivity ocean models (Models 1 and 2), both the amplitude and the phase change significantly as the core conductivity varies. In contrast, the high-conductivity ocean layer (Models 3 and 4) conceals the induction effect of the core. These trends are similar in the $Q_1$ transfer function for the orbital period $T = 1.370218$ d (see Figure S1). Therefore, a strong \textit{global} induction response ($A \gtrapprox 0.5$) at the synodic period can only be generated by a high-conductivity ocean ($\sigma_{ocean} \gtrapprox 5$ S/m), possibly in combination with a high-conductivity core. Medium amplitudes are consistent with either a conductive ocean or core. Weak induction amplitudes impose an upper bound on the conductivity of the ocean and core. Note that the amplitude of the induction response will further depend on the ice thickness (respectively, ocean thickness), with thinner ice shell (respectively, thicker ocean) resulting in stronger induction responses, and vice versa. 

\begin{figure}
\noindent\includegraphics[width=\textwidth]{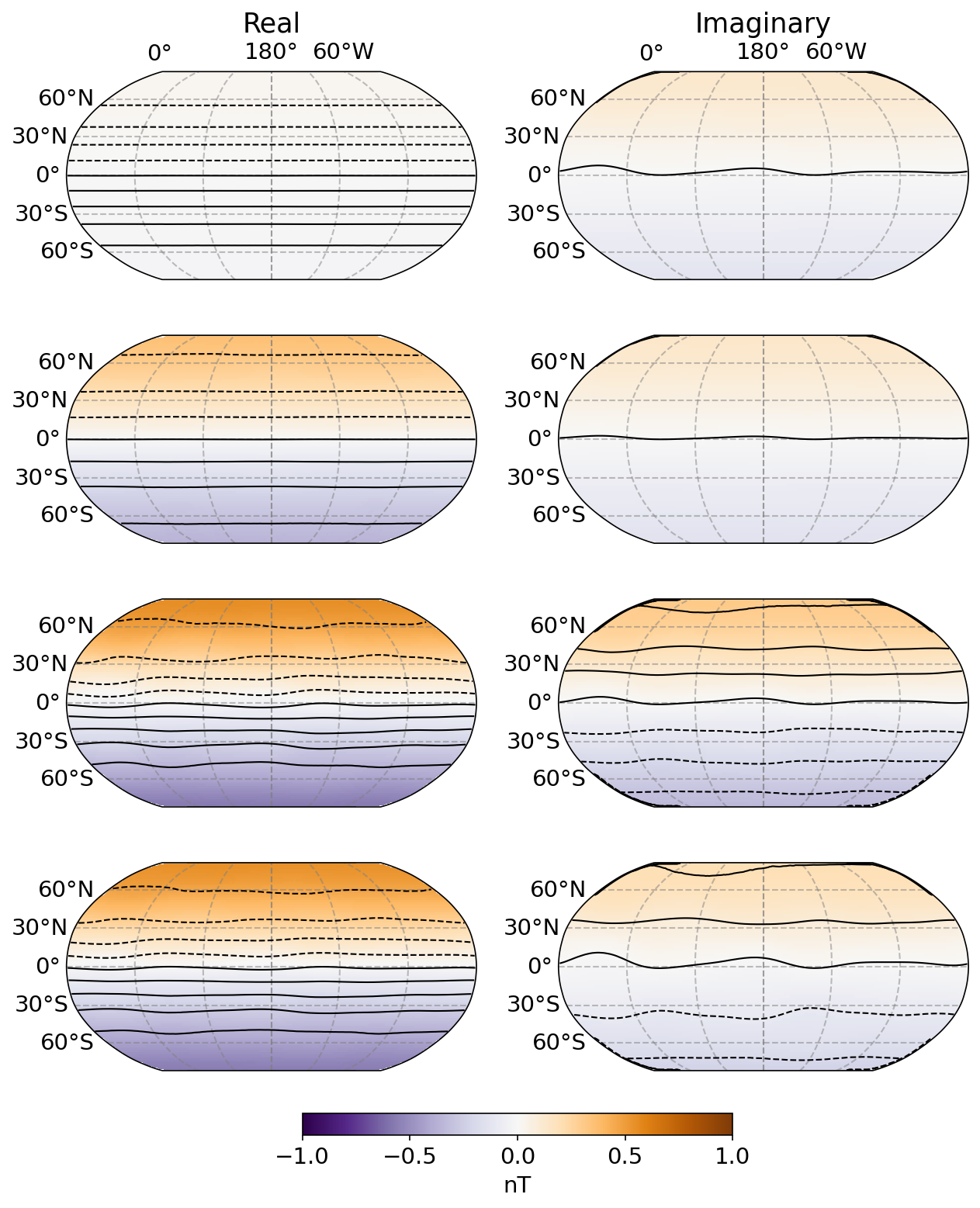}
\caption{Induced radial magnetic field at the surface of 3-D models (rows correspond to conductivity Models 1-4 in Table \ref{tab:models}). For all models, the external inducing field is represented by the $q_1^0 = 1$ nT coefficient (vertical dipole) oscillating with the synodic period. The equivalent Figure S2 shows the corresponding induced field for the $q_1^1 = 1$ nT external field geometry. Solid and dashed isolines correspond to positive and negative values, respectively.}
\label{fig:Bsurface}
\end{figure}

Next, we examine the 3-D induction effects arising from heterogeneous ice thickness. Figure \ref{fig:Bsurface} shows the induced magnetic field for all four 3‑D models, each driven by a unit vertical external field ($q_1^{0} = 1$ nT) that oscillates with the synodic period. The induced field is weakest for Model 1, corresponding to a low-conductivity ocean and core, and grows as the conductivity of the ocean and/or core increases. This tendency agrees with the 1-D results discussed in the previous paragraph. The similarity between 1-D and 3-D results is expected, as the 3-D conductivity anomaly arises solely from variations in ice thickness. In what follows, we will mostly analyse the differences between the magnetic fields for 1-D and 3-D models. Because the external inducing field is the same for both 1-D and 3-D models, their difference isolates the effect of the lateral conductivity heterogeneity on the induced field. 

\begin{figure}
\noindent\includegraphics[width=\textwidth]{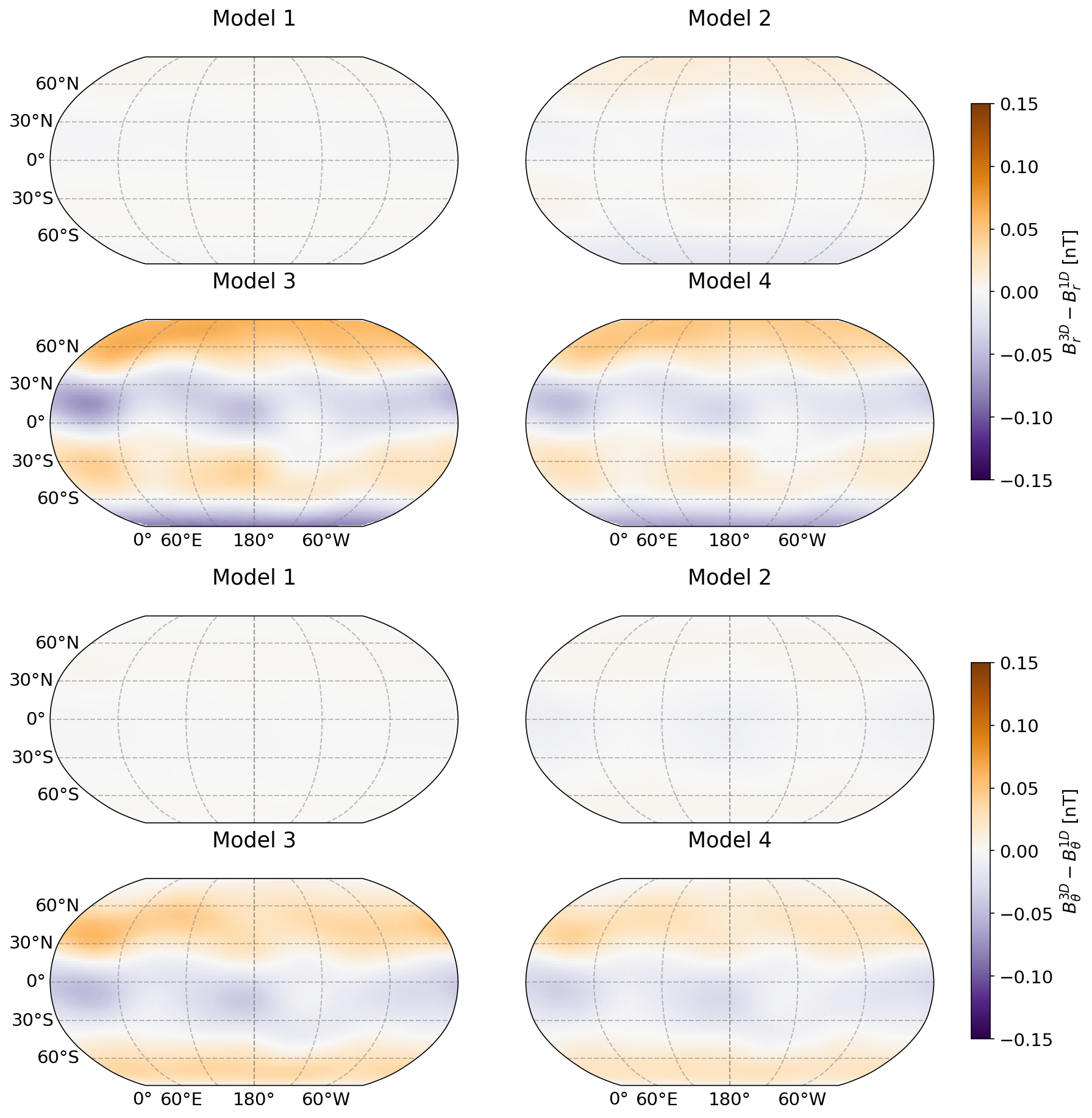}
\caption{Difference between real parts of the magnetic fields for 1-D and 3-D models for an external inducing field oscillating at the synodic period and described by $q_1^0 = 1$ nT. Models 1-4 correspond to four conductivity models defined in Table \ref{tab:models}. The upper and lower four panels show $B_r$ and $B_{\theta}$ components.}
\label{fig:dB_surface}
\end{figure}

\begin{figure}
\noindent\includegraphics[width=\textwidth]{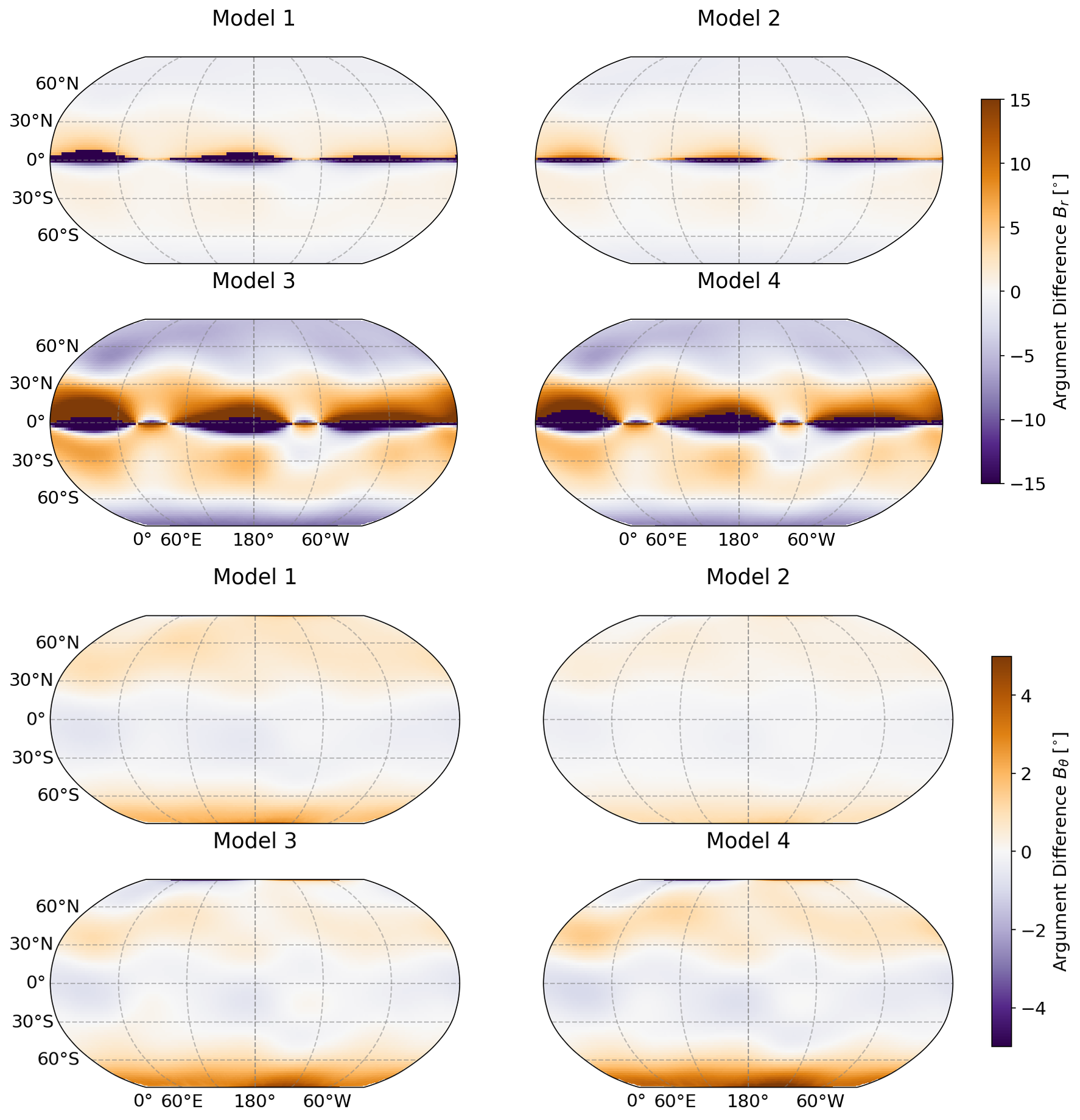}
\caption{Argument difference between magnetic fields for 1-D and 3-D models. Models 1-4 correspond to four conductivity models defined in Table \ref{tab:models}. The upper and lower four panels show $B_r$ and $B_{\theta}$ components.}
\label{fig:dPhi_surface}
\end{figure}

Figures \ref{fig:dB_surface}-\ref{fig:dPhi_surface} show the field and phase angle differences between 1-D and 3-D models. The differences are negligible over the surface if the conductivity of the ocean is low (Models 1-2). For high-conductivity ocean Models 3-4, these differences reach 0.15 nT and 15$^{\circ}$ in amplitude and phase (away from regions where the field components cross zero). Differences between 1-D and 3-D models correlate with variations in the ice shell thickness. The 3-D induced response is stronger than the 1-D response (computed with the mean ice thickness of 21 km) at poles where ice is thin, and weaker in regions where the ice thickness is larger. The effect of the core conductivity on 3-D responses appears relatively small, as is evident from small differences between Models 1-2 and Models 3-4 (corresponding to the 1st and 2nd columns). Results for the $q_1^1$ external-field geometry (presented in Figures S3 and S4) are consistent with the conclusions above. 

\begin{figure}
\noindent\includegraphics[width=\textwidth]{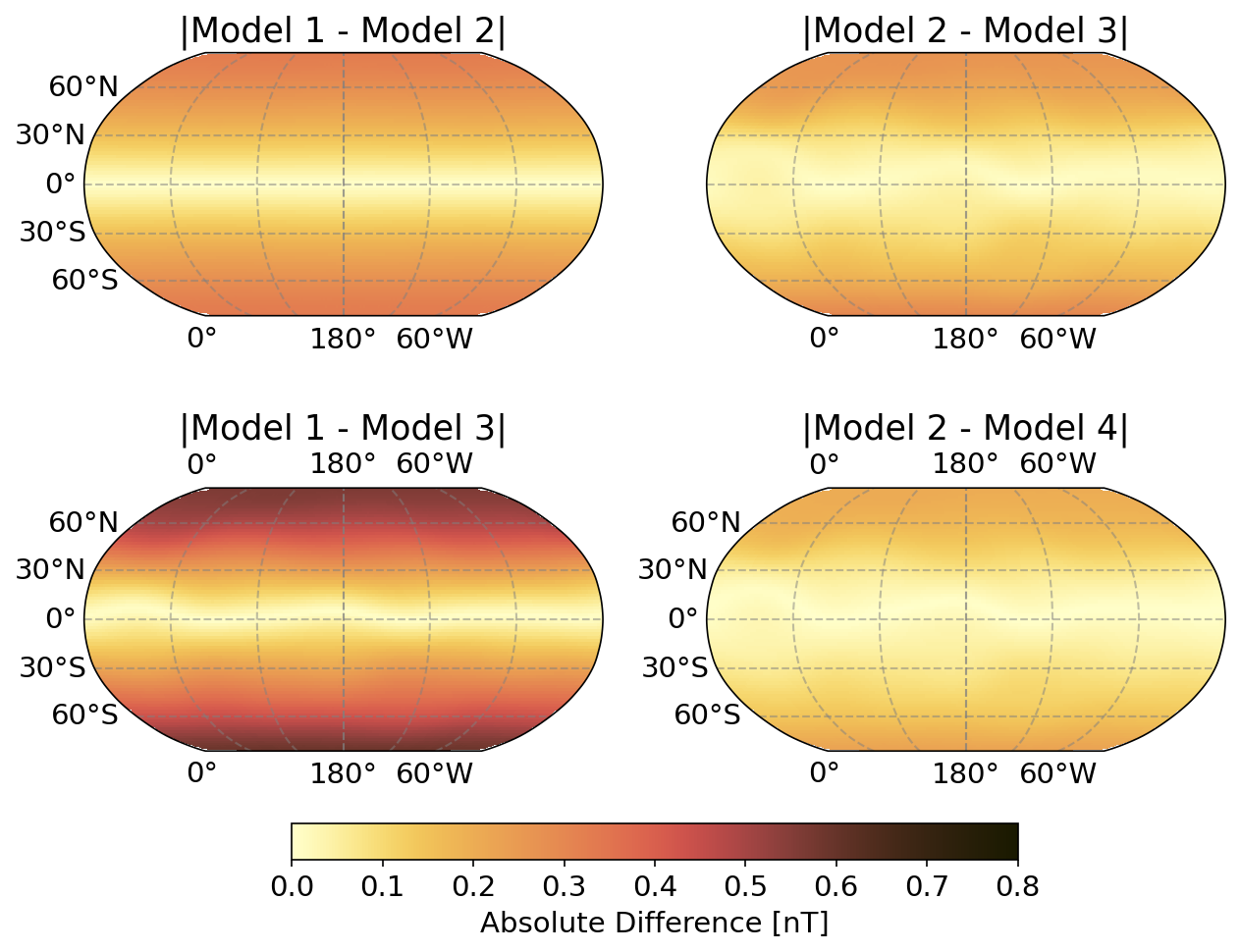}
\caption{Absolute differences in surface radial magnetic field between corresponding 3-D models relative to the external field described by $q_1^0 = 1$ nT and oscillating at the synodic period.}
\label{fig:model_diff}
\end{figure}

We further show absolute differences between 3-D models in Figure \ref{fig:model_diff}. The spatial pattern is dominated by the $\cos{\theta}$ term, amounting to the difference in the bulk 1-D response. The most significant difference ($> 0.5$ nT relative to the 1 nT inducing field) is between the models with low (Model 1) and high (Model 3) ocean conductivity, and low core conductivity. However, if the highly conductive core is present, the difference between low and high ocean conductivity models (Model 2 and 4) is less significant. The difference between Model 1 and 2, corresponding to low and high core conductivity, and a low-conductivity ocean, is also substantial. However, when the ocean is highly conductive, sensitivity to the core is minimal. Note that these findings are in line with our conclusions based on the 1-D $Q$-response analysis at the beginning of this section (Figure \ref{fig:q1d}). 

In addition to the synodic period results shown here, we also computed all results for the orbital period. Although the absolute values differ, the magnitudes of the 3-D effects and the overall trends are very similar to those of the synodic period across all models. 

\subsection{Local responses}
\label{sec:local}

In this section, we calculate the local $C$-response for 1-D and 3-D subsurface conductivity models. As was discussed in the theory section, the $C$-response is constant everywhere on the surface of a 1-D model and depends on the SH degree of the inducing field. For a 3-D model, the local $C$-response varies over the surface depending on the lateral gradients in the subsurface conductivity distribution and the external field geometry given by $Y_n^m$. Therefore, if the external field geometry is constrained or can be assumed, lateral variations in $C$-response are directly linked to variations in the subsurface electrical conductivity. 

\begin{figure}
\noindent\includegraphics[width=\textwidth]{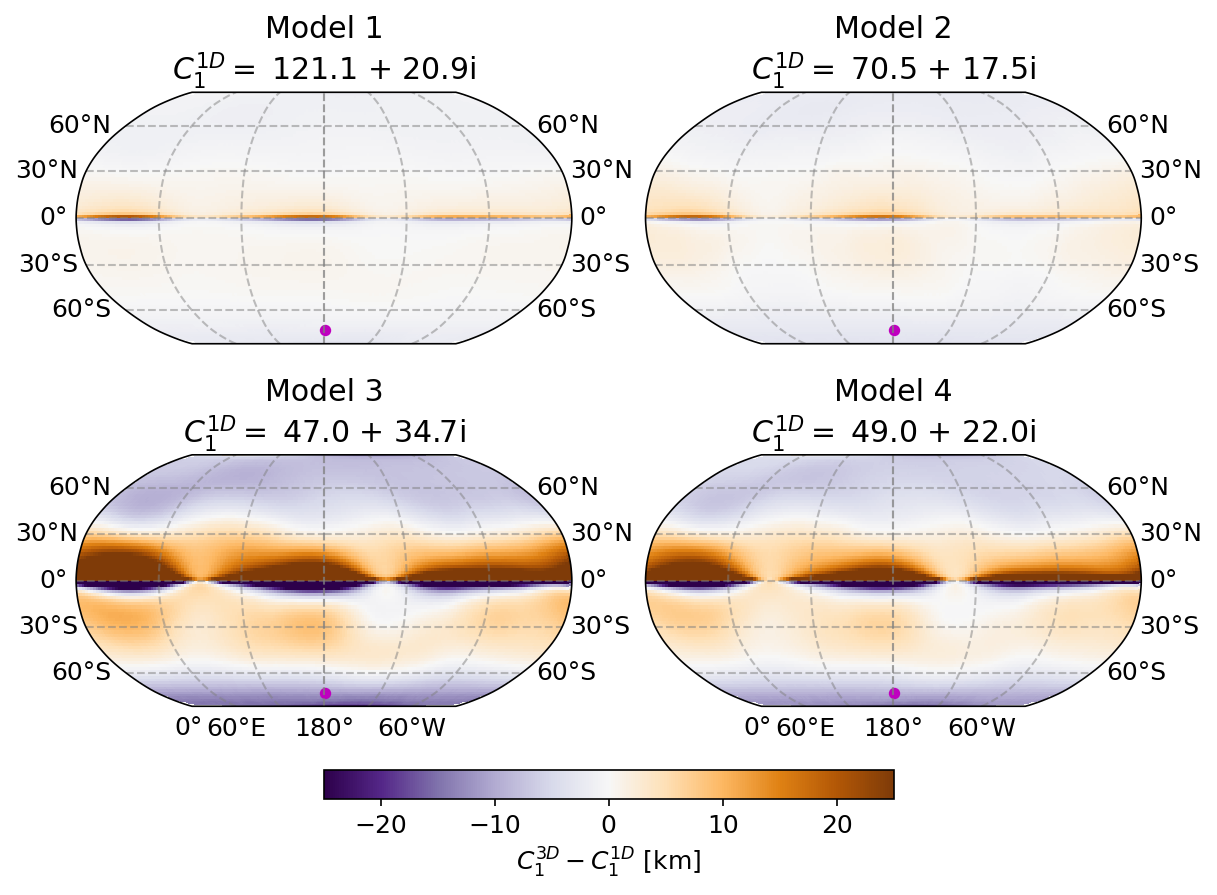}
\caption{Surface maps of differences between real parts of the $C$-response functions for 1-D and 3-D models at the synodic period. Constant values $C_1^{1D}$ for 1-D responses are given for each of the four models (Table \ref{tab:models}). The magenta point depicts the location at which broadband responses shown in Figure \ref{fig:clocal_q10} were computed.}
\label{fig:dC_synodic_q10}
\end{figure}

Figure \ref{fig:dC_synodic_q10} shows variations of $C$-responses over the surface of 3-D models relative to the 1-D response. Disregarding the region with zero radial magnetic field magnitude (corresponding to the equator for the $q_1^0$ geometry), we observe significant deviations from 1-D responses for models with a highly conductive ocean. At the poles, the real part of the $C$-response is smaller in the 3-D model than the 1-D model, in line with the smaller penetration depth due to the thin ice shell compared to the mean 1-D model with the 21 km thick ice. At the equator, where the ice shell thickness is larger than 21 km, the situation is the opposite: the real part of the 3-D $C$-response is larger than that of the 1-D response.

\begin{figure}
\noindent\includegraphics[width=\textwidth]{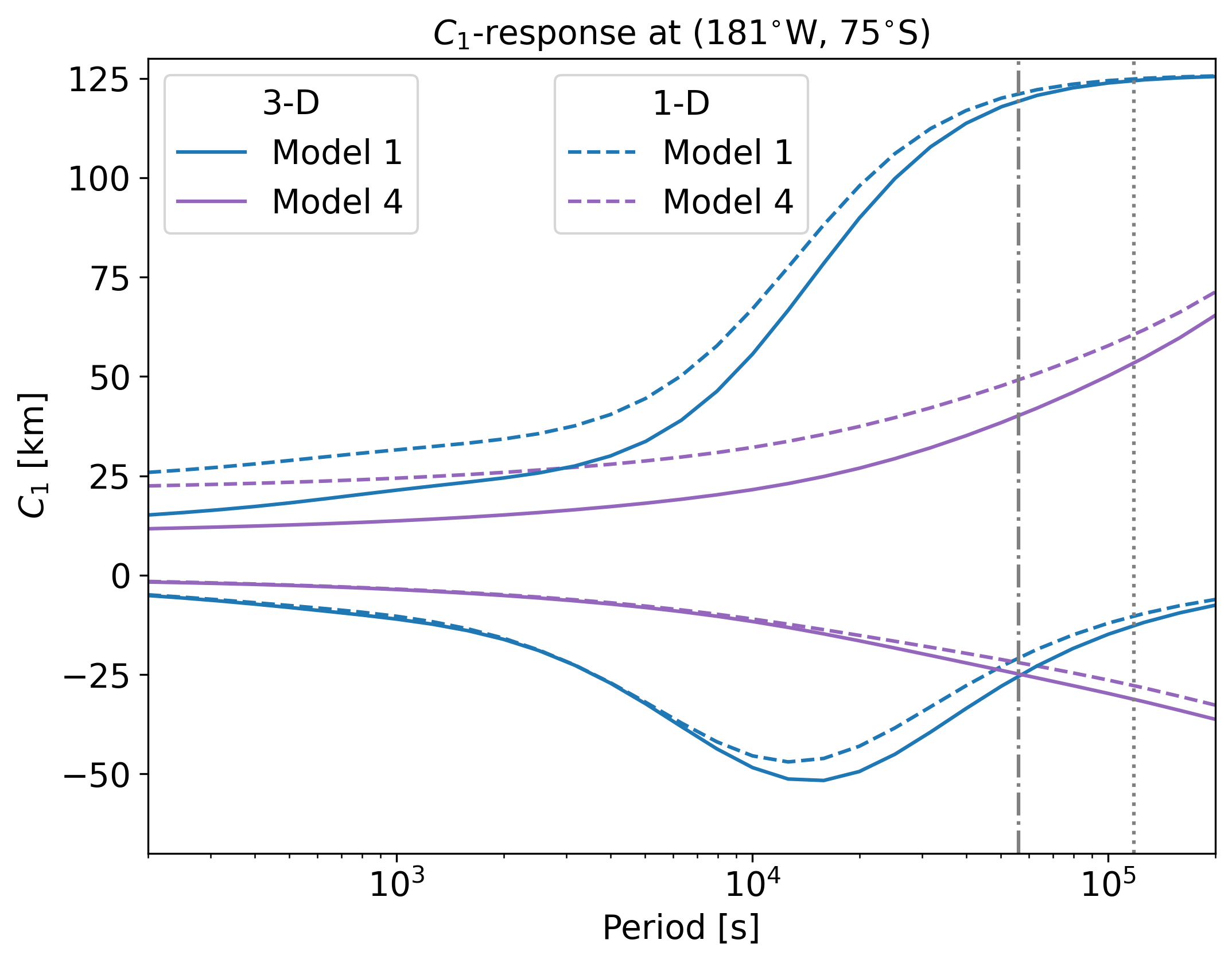}
\caption{Local $C$-response for low and high conductivity models (Table \ref{tab:models}) assuming 1-D (mean ice shell thickness of 21 km) or 3-D ice shell geometry. Real and Imaginary parts of the complex-valued $C$-response function are shown as positive and negative values, respectively. Vertical dash-dot and dotted lines correspond to the synodic and orbital periods, respectively.}
\label{fig:clocal_q10}
\end{figure}

Next, we computed the local $C$-response at one location for the period range $2\cdot10^2 - 2\cdot10^5$ s. Since the potential future lander may target the south pole region \cite{martins2024report}, we selected a location $181^{\circ}W, 75^{\circ}S$ that is close to the south pole. Figure \ref{fig:clocal_q10} shows the local 3-D and 1-D $C$-responses for two end-member models from Table \ref{tab:models} assuming a $q_1^0$ external (inducing) field geometry. At shorter periods, real parts of the transfer functions converge to the local ice thickness value of $7.5$ km and 21 km for 3-D and 1-D models, respectively. The imaginary parts differ the most at longer periods, signifying the 3-D induction effects caused by the ice shell structure. This indicates that as the period becomes shorter, local $C$-responses are primarily sensitive to closer and shallower conductivity structures. 

Figure S5 shows the local $C$-responses for the $q_1^1$ source geometry. Whereas transfer functions for 1-D models are the same for all external field geometries described by a single SH coefficient of the same degree, the 3-D responses are significantly different between the $q_1^1$ and the $q_1^0$ external field geometries. This is not specific to a chosen location and shows the importance of understanding the origin and structure of the inducing field at a potential landing site. Without reliable constraints on the external field geometry, the EM sounding with a single magnetometer is inherently limited.

Simulating EM induction responses for a broadband spectrum leads to several conclusions. First, at periods much shorter than $200$ s we can at best probe the first few km of the ocean. The transfer functions at the orbital period (potentially the longest detectable and well-constrained period for a lander instrument) will allow us to probe into the core. Therefore, estimation of transfer functions (TFs) within the range studied here will unlock a detailed EM sounding of all major layers, except the ice shell, for which much shorter periods would be needed. On Earth, solar wind interactions with the ionosphere-magnetosphere system and atmospheric phenomena generate a dense spectrum of natural external field variations ranging from milliseconds to the solar cycle \cite{constable2023grand}. These natural variations do not need to be periodic. A series of random, sporadic short-term pulses will generate a rich spectrum, and such sources are abundant in the Earth's ionosphere and atmosphere. This allows us to estimate EM transfer functions over a broad spectrum, providing ample opportunities for EM sounding from the surface down into the lower mantle \cite{grayver2024unravelling}. It is unclear whether a broadband external field spectrum is present at Enceladus, allowing for the estimation of a broadband local transfer function. We discuss these aspects further in the next section.

\begin{figure}
\noindent\includegraphics[width=\textwidth]{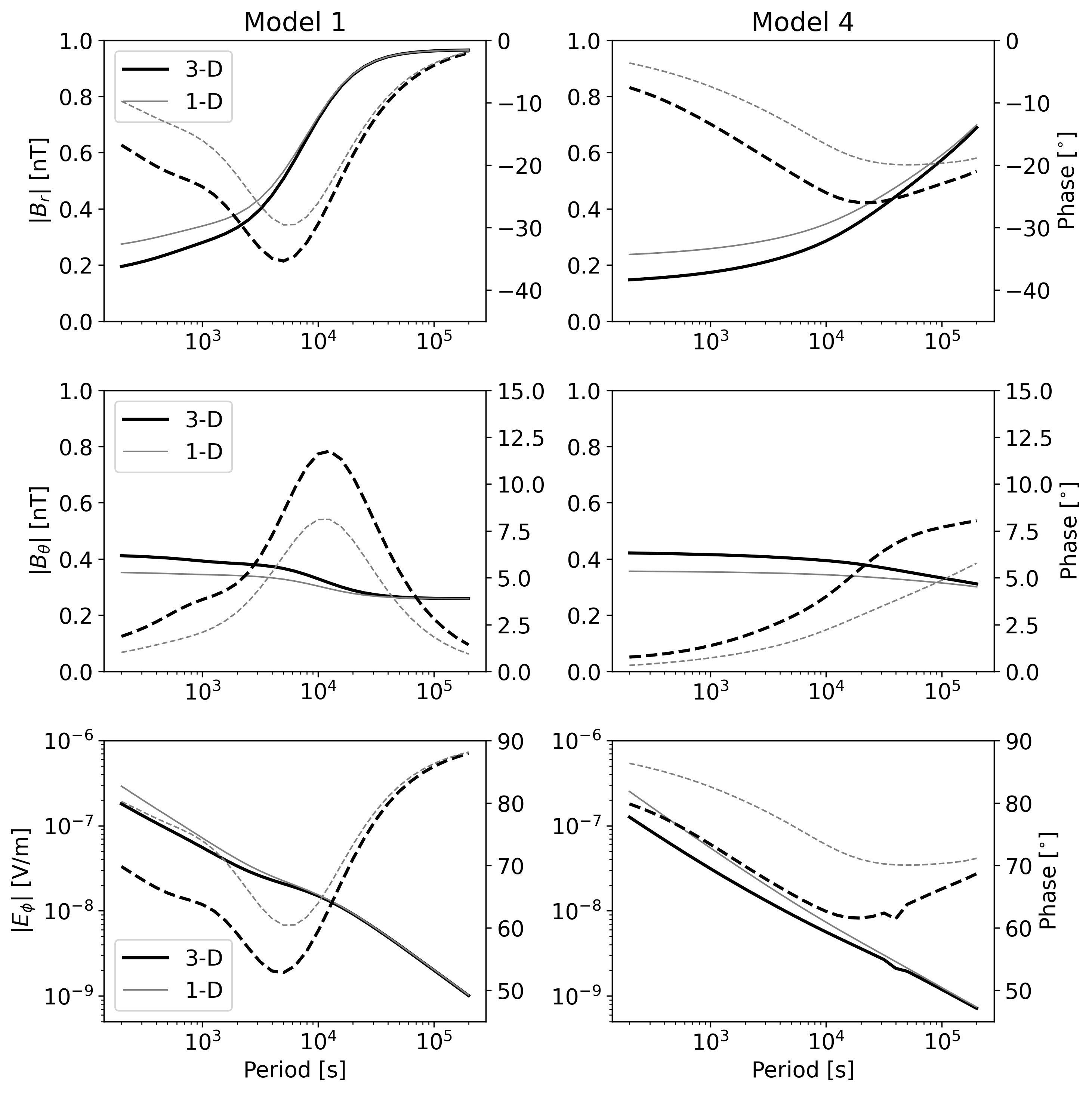}
\caption{Local surface horizontal magnetic and electric field components presented as the amplitude (solid) and phase angle (dashed) for 1-D (gray) and 3-D (black) models. Left and right columns represent two end-member conductivity models (Table \ref{tab:models}). Spectra are computed at the location $181^{\circ}W, 75^{\circ}S$ assuming a unit $q_1^0 = 1$ nT external field (the equivalent plot for the $q_1^1$ geometry is shown in Figure S6).}
\label{fig:fieldlocal_q10}
\end{figure}

In practice, TFs such as $C$-response are estimated from measurements of local magnetic and, if present, electric fields. Therefore, it is useful to characterize anticipated EM field variations across the spectrum. Figure \ref{fig:fieldlocal_q10} shows simulated EM fields over the considered spectrum (the corresponding plot for the $q_1^1$ geometry is shown in Figure S6). We note that the magnitude difference between 1-D and 3-D models, and hence the sensitivity, is most significant at shorter periods for both low- and high-conductivity models. Note that the actual field values on the surface of Enceladus will depend on the source geometry and magnitude, both of which can vary across the spectrum. However, these figures provide a tangible estimate of the relative magnitudes of the signals and their variation with subsurface conductivity. Such estimates can be expanded in future studies to inform future payload design.  

\section{Discussion and Conclusions}
\label{sec:disc}

This study elaborated on the electromagnetic induction in Enceladus. Specifically, we developed a model of the electromagnetic field induced in the interior of Enceladus by an external time-varying magnetic field. We considered models in which the electrical conductivity within Enceladus is radially symmetric or varies in all three dimensions. In the latter case, we used the actual shape of Enceladus rather than a perfectly spherical geometry. We then showed how local and global electromagnetic TFs can be derived from observations of magnetic and possibly electric fields. 

After constructing a series of 1-D and 3-D interior conductivity models of Enceladus, we simulated induced magnetic fields due to a uniform, time-harmonic external magnetic field of unit magnitude. To assess the feasibility of detecting the induced signals, their amplitudes should be rescaled by the expected external magnetic fields around Enceladus. A homogeneous time-varying field with the total magnitude of $\approx 5$ nT at orbital and synodic periods was found to be consistent with Cassini-Huygens magnetometer measurements \cite{Saur2024}. The independent analysis by \citeA{styczinski2024planetmag} also estimated an external field of magnitude $4.8$ nT at the orbital period. Figures \ref{fig:q1d} and S1 show that the amplitude of the induced field for a 1-D model with 21 km thick ice shell reaches $>80\%$ of the inducing field magnitude, but varies in the range $10-60\%$ for geophysically plausible models from Table \ref{tab:models}. This results in an absolute induced surface field magnitude of $2-3$ nT at synodic and orbital periods. 

The magnitude of 3-D EM induction effects for laterally heterogeneous models was found to be at most 0.15 nT relative to the mean 1-D model, which translates up to $\approx 0.8$ nT amplitude difference assuming a 5 nT external field. The 3-D effects are largest around the poles or in regions with strong gradients in ice thickness. Given much larger plasma-induced fields \cite{Saur2024}, the detection of induction effects of a non-spherical interior (3-D) will therefore be challenging for an orbiter. However, a sufficiently long orbiting phase and low-altitude measurements may enable this, since it allows the magnetic field sources to be better constrained and separated, as was achieved for Earth's magnetic field modeling, where the study of sub-nanotesla physical signals has become standard \cite{sabaka2020cm6, finlay2020chaos, grayver2024magnetic, fillion2023modeling, xiong2026influence}. 

In conclusion, EM induction responses with magnetic field amplitudes of up to a few nT at orbital and synodic periods can be expected, consistent with the latest Cassini-Huygens flyby data analysis \cite{Saur2024}. Knowledge of EM induction responses at synodic and/or orbital periods, with an accuracy of $\approx 1$ nT, would enable constraints on the bulk ocean conductivity and salinity structure. The EM response of the core is significant only if the core conductivity is high (e.g., the core is porous and the fluids are hot), while the ocean conductivity is low. If the ocean is highly conductive ($\gtrapprox 3$ S/m), the response of the core is concealed by the ocean. Mapping of 3-D induction effects produced by the ice shell thickness variations or thermo-chemical gradients \cite{ames2025ocean} would require sub-nanotesla accuracy, and is most feasible if the ocean conductivity is sufficiently high to produce a strong enough induction response. Therefore, the absence of variability in the magnetic field over the surface pertinent to ice shell thickness variations would be a strong indicator of either a low-conductivity (hence, low salinity) ocean or a thicker, more homogeneous ice shell. 

While space-borne observations can probe the global conductivity structure using longer (e.g. orbital and synodic) periods, local measurements offer an opportunity to estimate EM transfer functions over a broad spectrum, enabling detailed local conductivity sounding. In addition to EM sounding of the ocean and core, surface measurements will allow us to study local 3-D effects at periods shorter than the synodic period. Accuracy and sensitivity of $\lessapprox 0.1$ nT would be needed to detect variations across a broadband spectrum, which is well within the reach of established geomagnetic sensors \cite{hulot2021nanomagsat}. A larger unknown is whether the external field of sufficient strength exists at shorter periods to induce signals of detectable amplitude. This is plausible owing to the presence of ample MHD/plasma effects and significant observed field variance that cannot be explained by time-harmonic processes \cite{hadid2026evidence, Saur2024}. 

In practice, estimation of $C$-response (or impedances) based on a single-site magnetic field measurement requires prior knowledge about the source geometry (eqs. \ref{eq:c_via_h1}-\ref{eq:c_via_h2}). Another option is to estimate the local $C$-response from the ratio of horizontal electric and magnetic field components, where the explicit source geometry term is eliminated (compare eq. (\ref{eq:c_via_e}) and eqs. \ref{eq:c_via_h1}-\ref{eq:c_via_h2}), although the implicit dependency of EM field on degree $n$ remains. Using the horizontal electric-to-magnetic ratio for TF estimation and assuming a plane-wave source geometry renders the approach analogous to the Magnetotelluric method on Earth \cite{weidelt1972inverse, schmucker1985magnetic}. The validity of the plane-wave approximation is discussed in detail in \ref{app:planewave}. In general, the absence of the source geometry scaling term is a major advantage in practice since the dependency of the EM field on SH degree $n$ decreases at shorter periods (see \ref{app:planewave}). Local EM sounding in the period range $10^1 - 10^5$ s will unlock sensitivity to electrical conductivity structure throughout the ocean and the core, and help meet the science goal of a future mission. This conclusion was also reached in a recent independent study \cite{wivell2026inductive}.

The feasibility of measuring low-frequency electric fields on an icy, rugged surface is an open question. Among other factors, this means ensuring that electrodes are not polarized and that instrumental effects remain below the expected EM field amplitudes (Figure \ref{fig:fieldlocal_q10}). If the in-situ plasma density is high, additional corrections or the development of local plasma models may be required. As these aspects are out of scope, we note only that recent instrument developments on this topic are of interest \cite{grimm2021feasibility}, as is the Radio \& Plasma Wave Investigation (RPWI) payload onboard the ESA JUpiter ICy moons Explorer (JUICE), which will measure broadband electric field and help assess whether they are suitable for induction studies \cite{wahlund2025radio}, albeit from an orbit. Moreover, the methodology of carrying broadband MT sounding on terrestrial ice shields is well established \cite{hill2020temporal}.

As an alternative, a joint transfer function using both the orbiter and lander can be estimated, assuming that the orbiter observes only the external (inducing) component of the magnetic field, whereas the lander is sensitive to both the inducing and induced fields. This approach was previously used for Apollo 12 - Explorer 35 electromagnetic sounding experiment \cite{sonett1982electromagnetic}. However, depending on the lander location, this approach may prove infeasible at Enceladus owing to significant plasma-magnetic field interactions, which are strongest near the southern pole \cite{Saur2024}, and/or too low altitude of the orbiter. A combined use of an orbiter and a lander would, again, require a detailed understanding of the electromagnetic environment around Enceladus. 

The shape and ice thickness models (Table \ref{tab:params}) chosen here are not the only models available. For instance, an alternative model of \citeA{hemingway2019enceladus} overall agrees with that of \cite{Cadek2019} in that both predict thinner ice over the poles, with the former predicting a slightly thicker ice shell (e.g., $\approx 5$ vs $\approx 4$ km at the southern pole). These differences do not alter the conclusions of this study. More recently, \citeA{Park2024} published a model that predicts a significantly larger mean ice shell thickness of $\approx 30$ km compared to the $21$ km used by \citeA{Cadek2019} based on the shape model of \citeA{Tajeddine2017}. Additionally, the new model predicts a thinner ocean and a slightly larger, lighter (hence potentially more porous) core. If true, this would result in a smaller global induction response from the ocean. On the other hand, this would increase the relative contribution of the core, making the EM induction sounding even more suitable for probing core properties, such as porosity, salinity, and temperature, which are difficult to constrain with other geophysical techniques. 

In this study, we considered only the magnetic fields induced by an external time-varying magnetic field. Additionally, magnetic signals induced by the ocean flowing across magnetic field lines can constitute another mechanism for studying Enceladus's ocean. However, as we show in \ref{app:OIMF}, this mechanism likely requires high ($\gtrapprox 1$ m/s) ocean flow velocities to generate a detectable magnetic signature outside Enceladus. The required velocities are significantly larger than those reported in the current literature \cite{Zeng_2021, kang2022does, cabanes2024zonostrophic}. Therefore, the external time-varying magnetic field is likely the primary source of electric currents and associated induced magnetic fields inside Enceladus. Nevertheless, future research should remain open to this possibility as well.

The formalism developed in Section 2 provides a universal toolset for studying the interiors of moons and planets via electromagnetic induction sounding. The framework is not limited to a 1-D subsurface conductivity distribution and a homogeneous external field. We demonstrated how global and local electromagnetic transfer functions resulting from an arbitrary time-varying external magnetic field can be derived from magnetic (and potentially electric) field measurements. These transfer functions can be used to constrain the interior electrical conductivity and to infer the thermo-chemical properties of the underlying materials. The major challenge associated with the application of EM sounding is the complexity of electromagnetic environments around moons and planets, characterized by ample plasma and MHD effects and small-scale heterogeneities that can interfere with the assumptions underlying EM induction sounding techniques. The cleanliness of the observation platform, magnetometer accuracy, and spacecraft orbit impose additional constraints in practice. The physical framework provided herein is therefore a valuable tool in assessing the potential of EM sounding for future missions or revisiting existing observations.

\appendix
\section{Plane-wave approximation}
\label{app:planewave}

We denote $Z$ to be the impedance at the surface of a flat layered model excited by a homogeneous vertically incident plane wave. In case of the homogeneous half-space with conductivity $\sigma$, the impedance is \cite{berdichevsky2002magnetotellurics}
\begin{equation}
    Z = \sqrt{\frac{\mathrm{i}\omega\mu}{\sigma}} = \frac{1+\mathrm{i}}{\sigma d_s},
\end{equation}
where the skin depth
\begin{equation}
    d_s = \sqrt{\frac{2}{\omega\mu\sigma}}
\end{equation}
corresponds to a depth at which the amplitude of an incident plane-wave field decreases $e$ times. Following the relation (\ref{eq:C2Z}), we can also express the $C$-response of a homogeneous half-space model as
\begin{equation}
    C = \sqrt{-\frac{1}{\mathrm{i}\omega\mu\sigma}} = \frac{1+\mathrm{i}}{2}d_s.
\end{equation}
We see that the real part of the plane-wave $C$-response is equal to half of the skin depth and represents the "center-of-mass" of the induced current density \cite{weidelt1972inverse}. The plane-wave impedance is often referred to as the magnetotelluric (MT) impedance in Earth induction studies, alluding to the MT method and the fact that it is estimated from observations of natural magnetic and electric field variations \cite{tikhonov1950determining, cagniard1953basic}. Conventionally, the MT impedance is presented as a function of frequency (period) and is plotted in terms of the apparent resistivity
\begin{equation*}
    \rho^{\mathrm{app}} = \frac{|Z|^2}{\omega\mu}=\omega\mu|C|^2
\end{equation*}
and phase
\begin{equation*}
    \Phi = \arg{Z}.
\end{equation*}
It is easy to verify that for a homogeneous model, $\rho^{app} = \sigma^{-1}$ and $\Phi = \pi/4$.

\begin{figure}
\noindent\includegraphics[width=\textwidth]{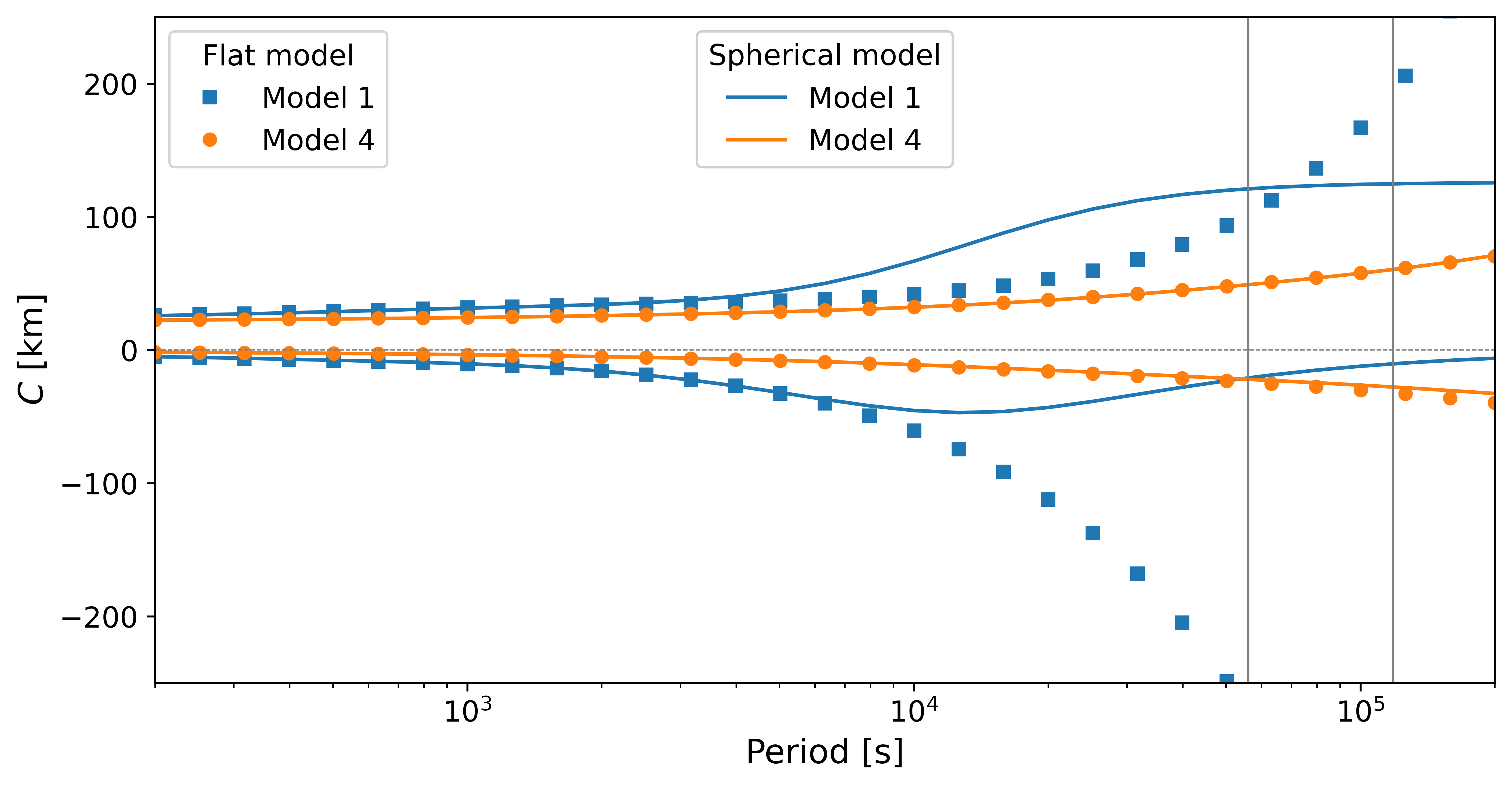}
\caption{$C$-responses computed for the spherical ($C_1$) and the plane-wave models for the same subsurface conductivity profile. Real and imaginary parts of the $C$-response correspond to positive and negative values, respectively. Vertical solid lines denote synodic and orbital periods. Model numbers refer to Table \ref{tab:models}.}
\label{fig:A1}
\end{figure}

\begin{figure}
\noindent\includegraphics[width=\textwidth]{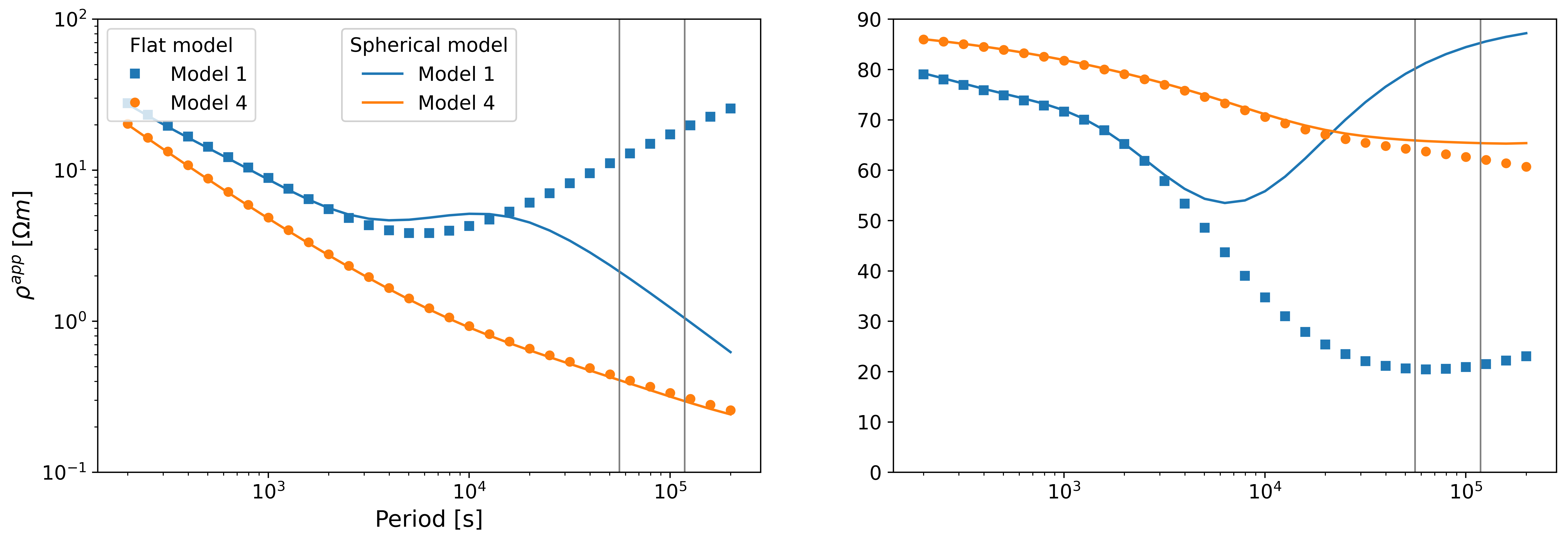}
\caption{Same as Figure \ref{fig:A1}, but presenting the $C$-response in the form of the apparent resistivity and phase curves.}
\label{fig:A2}
\end{figure}

The corresponding spherical quantities $Z_n, C_n$ will approach the plane-wave limit if the condition in eq. (\ref{eq:planewave_cond}) holds. The equations above, stated for a homogeneous half-space model, can be used to get a crude estimate when the plane-wave approximation is permissible. However, a more rigorous approach is to compare the spherical ($Z_n, C_n$) and flat model ($Z, C$) quantities for the same 1-D conductivity profile. Figure \ref{fig:A1} shows the spherical and plane-wave $C$-responses computed for the most resistive and most conductive Models 1 and 4, respectively. For a high-conductivity subsurface, plane-wave is a valid approximation up to the orbital period, whereas for a low-conductivity model, the plane-wave approximation breaks down already at a period of a few hours. Figure \ref{fig:A2} presents the corresponding apparent resistivity and phase curves, which support the conclusions above.

Formally, a plane-wave diffusing within a layered flat model has an infinite lengthscale, such that magnetic field lines have no start or end points, thereby satisfying $\nabla \cdot \mathbf{B} = 0$, but requiring infinite energy. Therefore, an infinite plane-wave is not a physically valid model, and $Z$ should only be treated as a local field approximation, whereas $Z_n$ represents fundamental solutions of the quasi-static Maxwell's equations in a spherical conductor \cite{srivastava1966theory}.
As a result, the spherical impedance and, hence, $C$-response, depend on the SH degree $n$ that is related to the spatial scale of the inducing field. Therefore, both the period and SH degree $n$ determine the diffusion length in a spherical conductor \cite{schmucker1985magnetic}. 

\begin{figure}
\noindent\includegraphics[width=\textwidth]{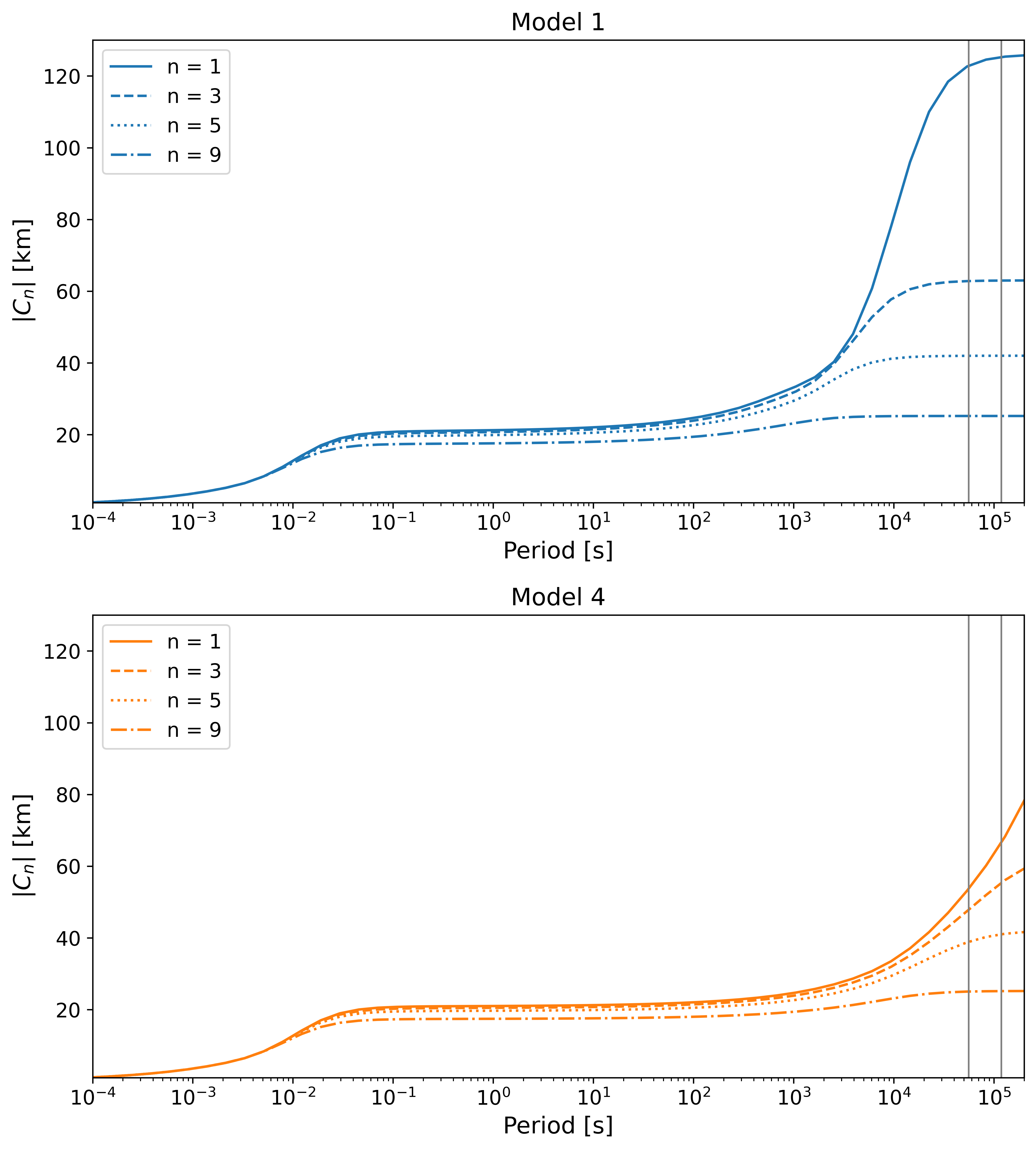}
\caption{Amplitude of the $C$-responses computed for different SH degrees $n$ as a function of period using the same subsurface conductivity profile. The top and bottom plots correspond to low-conductivity Model 1 and high-conductivity Model 4, respectively.}
\label{fig:A3}
\end{figure}

The results in the main text were computed for the homogeneous planetary field ($n=1$). However, higher external field harmonics may contribute to, or even dominate, the local field. Therefore, Figure \ref{fig:A3} shows the magnitude of the $C_n$ for different SH degrees. The responses differ significantly at periods $> 10^4$ s, when the fields diffuse into the ocean and core. At shorter periods, a broadband plateau emerges, where responses are nearly constant with values close to the ice thickness, indicating that fields attenuate within the upper fraction of the ocean. Finally, at higher frequencies ($> 100 Hz$) all responses merge and, according to the value of the real part, attenuate within the ice shell. From these results, the plane-wave regime can only be justified for the high-frequency part of the spectrum. However, the sounding depth is then restricted solely to the ice shell.  

\section{Ocean Induced Magnetic Field}
\label{app:OIMF}

In addition to the external time-varying magnetic field, the flow of electrically conductive fluid across the ambient magnetic field leads to the generation of the electric current density within the ocean \cite{faraday1832vi} that can be expressed as
\begin{equation}
\label{eq:lorentz}
    \mathbf{J} = \sigma_{\mathrm{ocean}}\left( \mathbf{E} + \mathbf{u} \times \mathbf{B}^0\right),
\end{equation}
where $\mathbf{u}$ is the ocean flow velocity and $\mathbf{B}^0$ the ambient magnetic field, dominated by the Kronian field at Enceladus. By Ampere's law, this current density will produce the magnetic field $\mathbf{b}$ (we use the small letter to signify that this field is a perturbation to the total field such that $\mathbf{B}^0 \gg \mathbf{b}$), which we refer to as the ocean-induced magnetic field (OIMF). Furthermore, we imply only the OIMF outside the ocean where it can be observed.

The OIMF has been routinely extracted from Earth's satellite magnetic observations \cite{sabaka2015cm5, grayver2024magnetic} and is used to probe the electrical conductivity of the mantle beneath the oceans \cite{grayver2016satellite, sachl2022inversion}. For reference, strong lunar tidal currents $u \sim \mathcal{O}(10^{-2}-10^{-1} \; \mathrm{m/s})$ on Earth \cite{egbert2017tidal} generate a few nanoTesla large-scale OIMF at the surface in the presence of the $\mathbf{B}^0 \sim \mathcal{O}(10^4 \; \mathrm{nT})$ ambient magnetic field. Combining eqs. (\ref{eq:lorentz}) and (\ref{eq:maxwell_fd1}) yields the scale relationship
\begin{equation*}
    |\mathbf{J}| \approx \frac{|\mathbf{b}|}{a\mu L},
\end{equation*}
where $\mathbf{b}$ is the magnetic field perturbation (OIMF) generated by the current density (\ref{eq:lorentz}) and $L \sim \mathcal{O}(10^3)$ km is the characteristic lengthscale of OIMF, and $a \ll 1$ represents geometric attenuation, cancellation by the induced field, and the fact that only the externally visible poloidal component contributes to the observed magnetic perturbation \cite{velimsky2019global}. We can assume that the conductivities of Earth's and Enceladus's oceans are similar. In contrast, the ambient magnetic field at Enceladus is $\approx 340$ nT \cite{dougherty2005cassini}, or roughly two orders of magnitude smaller compared to Earth. It follows from the scaling relation above that to generate a $\approx 1$ nT OIMF with lengthscale $L$ within the ocean of Enceladus, ocean flow velocities on the order of 1 m/s or larger are required. Recent studies infer smaller large-scale velocities \cite{Zeng_2021, kang2022does, cabanes2024zonostrophic}. Therefore, motional induction is unlikely to be a mechanism that can generate a strong magnetic signal. This analysis also holds for other icy moons, as demonstrated in the study by \citeA{vsachl2025magnetic}, which presents full-physics simulations of OIMF at Europa.

%
%

\section*{Conflict of Interest Statement}

The authors declare no conflict of interest.

\section*{Open Research Section}
The surface radius model was obtained from \citeA{Tajeddine2017}. The ice shell thickness model was obtained from the corresponding author of \citeA{Cadek2019}. The study results are available from \citeA{grayver_2026_19971354}. 

\acknowledgments
A.G. was supported by the Heisenberg Grant from the German Research Foundation, Deutsche Forschungsgemeinschaft (Project No. 465486300). J.S. received funding from the European Research Council (ERC) under the European Union Horizon 2020 research and innovation programme (grant agreement No. 884711). This work has benefited from discussions with Stefan Duling and Lorenz Roth.

\bibliography{references}

\end{document}